\documentclass[12pt,preprint]{aastex}

\usepackage{epsfig}

\newcommand{\Hm}{\rm{H}^{-}}
\newcommand{\me}{\rm{e^{-}}}
\newcommand{\Hp}{\rm{H}^{+}}
\newcommand{\Dp}{\rm{D}^{+}}
\newcommand{\Hep}{\rm{He}^{+}}
\newcommand{\mH}{\rm{H}}
\newcommand{\mD}{\rm{D}}
\newcommand{\He}{\rm{He}}
\newcommand{\mHt}{\rm{H}_{2}}
\newcommand{\hd}{\rm{HD}}
\newcommand{\mHtp}{\rm{H}_{2}^{+}}
\newcommand{\hi}{\hbox{H\,{\sc i}}\,}
\newcommand{\hii}{\hbox{H\,{\sc ii}}\,}

\newcommand{\oi}{\hbox{O\,{\sc i}}\,}
\newcommand{\siI}{\hbox{Si\,{\sc i}}\,}

\newcommand{\mC}{\rm{C}}
\newcommand{\cp}{\rm{C^{+}}}
\newcommand{\mO}{\rm{O}}
\newcommand{\op}{\rm{O}^{+}}
\newcommand{\oh}{{\rm OH}}
\newcommand{\mSi}{\rm{Si}}
\newcommand{\sip}{\rm{Si^{+}}}
\newcommand{\sipp}{\rm{Si^{++}}}
\newcommand{\expf}[3]{\exp \left(#1\frac{#2}{#3}\right)}
\def\simless{\mathbin{\lower 3pt\hbox
   {$\rlap{\raise 5pt\hbox{$\char'074$}}\mathchar"7218$}}}   
\def\simgreat{\mathbin{\lower 3pt\hbox  
   {$\rlap{\raise 5pt\hbox{$\char'076$}}\mathchar"7218$}}} 

\usepackage{natbib}
\bibpunct{(}{)}{;}{a}{}{,}

\shorttitle{Chemistry and cooling at low $n$ and low ${\rm Z}$}
\shortauthors{Glover \& Jappsen}

\begin{document}

\title{Star formation at very low metallicity. I:  Chemistry and cooling at low densities}

\author{S.~C.~O. Glover$^{1,2}$, A.-K. Jappsen$^{1,3}$}

\affil{$^1$Astrophysikalisches Institut Potsdam,\\An der Sternwarte 16, 14482
 Potsdam, Germany; sglover@aip.de}
\affil{$^2$Department of Astrophysics, American Museum of Natural History,
\\79th Street at Central Park West, New York, NY 10024-5192, USA}
\affil{$^3$ Canadian Institute for Theoretical Astrophysics,\\
University of Toronto, 60 St.\ George Street, Toronto, ON M5S 3H8, Canada;
jappsen@cita.utoronto.ca}

\begin{abstract}
We present a simplified chemical and thermal model designed to allow
computationally efficient study of the thermal evolution of metal-poor gas
within large numerical simulations.  Our main simplification is the neglect
of the molecular chemistry of the heavy elements. The only molecular
chemistry retained within the model is the formation and destruction of
molecular hydrogen. Despite this major simplification, the model allows
for accurate treatment of the thermal evolution of the gas within a large
volume of parameter space. It is valid for temperatures $50 < T < 10000
\: {\rm K}$ and metallicities $0 <  {\rm Z} < 0.1\: {\rm Z_{\odot}}$. In gas with
a metallicity ${\rm Z} = 0.1 \: {\rm Z_{\odot}}$, and in the absence of an
incident ultraviolet radiation field, it is valid for hydrogen number densities
$n_{\rm H} \simless 500 / t_{\rm char} \: {\rm cm^{-3}}$, where $t_{\rm char}$ is
the size in Myr of the characteristic physical timescale of interest in the 
problem. If ${\rm Z} \ll 0.1 \: {\rm Z_{\odot}}$, or if a strong ultraviolet radiation 
field is present, then the model remains accurate up to significantly higher 
densities. We also discuss some possible applications of this model.
\end{abstract}

\section{Introduction}
It has long been known that cooling by molecular hydrogen, $\mHt$,
plays a major role in regulating star formation in primordial protogalaxies
\citep{SAS67,PEE68,MAT69,LEP83,TEG97,ABE02}. The importance of $\mHt$ 
stems from the fact that in most circumstances it is the dominant coolant in 
primordial gas at  $T < 10^{4} \: {\rm K}$. As a result, the chemistry of $\mHt$ 
in primordial gas has attracted considerable study 
\citep{dl87,bl91,ABE97,GP98,sld98, lsd02} 
and it is now generally accepted that only a small chemical network, 
of maybe 20--30 reactions, is required to model $\mHt$ chemistry over a 
very wide range of conditions in these systems. 

The introduction of metals into the gas, as will occur following enrichment
of their surroundings by the first generation of supernovae, complicates
matters enormously. Many other atomic and molecular coolants become
available and the associated chemistry is highly complex: for instance,
a reasonably complex model of purely gas-phase chemistry can easily
stretch to $\sim 400$ reactants and almost 4000 reactions 
\citep[e.g.][]{TEU00}.
This would not matter were it not for two important points. First, many
astrophysicists believe that metal enrichment above a certain level
-- the so-called `critical metallicity' -- leads to a significant change in
the stellar initial mass function (IMF), from an IMF dominated by massive
stars to one that looks far more like the familiar Salpeter IMF of local
star formation \citep{BRO01,SCH02}. Second, testing this idea numerically 
using three-dimensional hydrodynamical simulations requires us to model 
the chemistry of the gas, but the highly detailed chemical models
mentioned above are impractical to use in high-resolution numerical 
simulations, owing to their high computational cost. This is a consequence
of the fact that chemical rate equations are frequently stiff and so for
reasons of  stability must be solved implicitly, with a computational cost 
that scales as the cube of the number of chemical species involved.

It is therefore important to look for ways to simplify the chemistry without
unduly compromising the accuracy of the resulting model. Major 
simplifications can be made if we make the reasonable assumption 
that the main coolants in low metallicity, high redshift gas will be 
similar to those in the local interstellar medium -- which should be
true provided that the abundance ratios of the various metals in
low metallicity gas are not {\em too} unusual -- and also if we restrict 
the range of physical conditions in which we are interested. 

In this paper, we present a chemical model designed to model the
chemistry of the major coolants in cool ($T \simless 10^{4} \: {\rm K}$)
low-metallicity gas at low gas densities. In the absence of ultraviolet 
radiation, and in gas with ${\rm Z} = 0.1 \: {\rm Z_{\odot}}$, our model 
is valid for atomic hydrogen number densities 
$n_{\rm H} \simless 500 / t_{\rm char} \: {\rm cm^{-3}}$,
where $t_{\rm char}$ is the size in Myr of the characteristic physical 
timescale of interest in the problem, corresponding to an overdensity of 
$\delta = 2.5 \times 10^{9} t_{\rm char}^{-1} (1+z)^{-3}$ with respect to 
the cosmological mean background density. At lower metallicities, or if
a moderately strong UV field is present, our model remains accurate up 
to significantly higher densities. We do not
treat the cooling or chemistry of hot gas ($T \gg 10^{4} \: {\rm K}$) as
this has already been treated in detail elsewhere \citep[see e.g.][]{sd93}.

The structure of this paper is as follows. In \S\ref{chem} we present our 
simplified chemical network and discuss the principles determining our 
choice of reactants (\S\ref{chem_choice}) and reactions (\S\ref{react_select}), 
as well as our treatment of grain surface chemistry (\S\ref{surface_chem}) and 
photochemistry (\S\ref{photochem}). In \S\ref{therm}, we discuss our 
treatment of the main thermal processes included in our model, with 
a particular emphasis on atomic fine structure cooling. We conclude 
in \S\ref{appl} with a brief discussion of possible applications of our
model.

\section{Chemical model}
\label{chem}
\subsection{Choice of chemical species}
\label{chem_choice}
Our choice of which chemical species to include in our simplified network was guided 
by two main considerations. In order to properly model the thermal evolution of the gas,
we must be able to accurately model the evolution of the chemical abundances of 
all of the major coolants. At the same time, in order to keep our chemical treatment 
computationally efficient, we do not want to include more species than are strictly 
necessary. In some cases, the decision on whether or not to include a species was
obvious. For instance, neutral atomic hydrogen, $\mH$, is a major constituent of the
gas and is also a major coolant at $T \sim 10^{4} \: {\rm K}$ and above. Molecular
hydrogen, $\mHt$, also must be included as it has long been recognised to be the 
dominant coolant in metal-free gas at  $200 < T < 10^{4} \: {\rm K}$ \citep{SAS67,PEE68}, 
and it is now clear that it also remains significant in metal-poor gas 
\citep{OMU05,SAN06,jgkm07}. A good case can also be made for the inclusion of
$\hd$, which dominates the cooling in zero-metallicity gas at $T < 200 \: {\rm K}$ 
\citep{FLO00}, and which remains important at least up to metallicities 
${\rm Z} \sim 10^{-5} \: {\rm Z_{\odot}}$ \citep{OMU05}. As far as metals go, however,
the choice is less obvious. Metal atoms and ions produce little in the way of resonance
line cooling at $T < 10^{4} \: {\rm K}$, but many can act as sources of fine structure 
line emission or emission from metastable transitions at these temperatures. Therefore, 
in order to determine which of the various species are important coolants in the region of 
parameter space that our model is designed to cover, we directly compared the 
cooling rates produced by each species, under the assumption that the relative 
abundances of the various elements were the same as in the local interstellar
medium. 

The species we investigated in this comparison were the same as those included 
in the shock models of \citet{HOL89}: $\mC$, $\cp$, ${\rm Cl}$, ${\rm Cl^{+}}$, 
${\rm Fe}$, ${\rm Fe^{+}}$, ${\rm N}$, ${\rm N^{+}}$, ${\rm Ne^{+}}$, ${\rm Ni}$, 
${\rm Ni^{+}}$, $\mO$, ${\rm O^{+}}$, ${\rm S}$, ${\rm S^{+}}$, ${\rm Si}$ and 
${\rm Si^{+}}$. This list includes all of the neutral or singly ionized species that
have both non-negligible abundances and emission lines that are accessible 
at $T < 10^{4} \: {\rm K}$. Doubly ionized species are unlikely to be abundant
in gas of this temperature, and so are unlikely to contribute significantly to the
total cooling rate. Atomic data for $\mC$, $\cp$, $\mO$, $\mSi$ and $\sip$ was 
taken from the sources listed in Tables~\ref{fs_data}--\ref{fs_coll_rates}. For the 
remaining species, we used data taken from \citet{HOL89}. For the relative
abundances of C, Fe, N, Ne, Ni, O, S and Si, we used values taken from 
\citet{sem00}, while for Cl we used a value taken from \citet{asp06}.

We investigated three different scenarios for the ionization state of the gas,
which we here refer to as the `no ionization', `moderate ionization' and 
`high ionization' cases. In the no ionization case,  we assumed that the gas 
was completely neutral, and so set the electron abundance to zero, along
with the abundances of all of the ionized species. In the moderate ionization case,
we assumed that the fractional ionization of hydrogen was such that
$n_{\Hp} / n_{\mH} = 10^{-4}$, that species with ionization potentials lower
than that of hydrogen -- C, Cl, Fe, Ni, S and Si -- were fully ionized, and that 
all of the other species remained fully neutral. Finally, in the high ionization case, 
we assumed that $n_{\Hp} / n_{\mH} = 0.5$, that species with ionization 
potentials lower than that of hydrogen were again fully ionized, and that
the fractional ionization of the other species was the same as that of hydrogen,
i.e.\ 50\%. 

In each case, we computed the cooling rate due to fine structure and/or
metastable transitions from each of the listed species for a large number of
temperatures in the interval $50 < T < 10000 \: {\rm K}$ and  number 
densities in the interval $0.001 < n < 100 \: {\rm cm^{-3}}$. In these calculations,
we assumed that the total metallicity of the gas was $0.1 \: {\rm Z_{\odot}}$, and 
that the gas was optically thin in all of the relevant transitions. We also calculated 
the contributions to the total cooling rate made by \hi Lyman-$\alpha$ emission
and Compton cooling; the rate of the latter was calculated for an assumed
redshift $z=20$.

We used the results of our comparison to select the set of major coolants that
it was necessary to include in our thermal model by identifying all of the coolants 
that contributed more than 25\% of the total cooling rate for any of the 
combinations of temperature, density and ionization that we examined. 
The resulting set consisted of $\mC$, $\cp$, $\mO$, $\sip$, Compton 
cooling and Lyman-$\alpha$ cooling. Therefore, the only metals that we 
include in our chemical and thermal model are carbon, oxygen and silicon.
The importance of carbon and oxygen is unsurprising -- they are well known
to play a major role in the cooling of the local atomic ISM \citep{WOL03},
and previous authors have also predicted that they will play a key role at
high redshift \citep[see e.g.][]{BRO01,BL03,SAN06,FJB07}. Silicon has 
attracted less attention (although see \citealt{SAN06}), but $\sip$
proves to be the dominant coolant in highly ionized gas with $700 \simless
T \simless 7000 \: {\rm K}$ at $n > 0.3 \: {\rm cm^{-3}}$. Although gas with
the very high fractional ionization assumed here will recombine quickly
at temperatures $T < 10^{4} \: {\rm K}$, it nevertheless
seems prudent to include silicon in our model.  Moreover, since it is necessary
to include neutral silicon in our chemical model if we are to calculate the
$\sip$ abundance accurately, it requires little extra effort to include the effects
of fine-structure cooling from \siI.

Our choice of a 25\% cutoff in this analysis is somewhat arbitrary. If we were
to decrease the size of this cutoff, we would find that the number of species
that must be included would increase, as both ${\rm Fe^{+}}$ and S are 
important coolants at the 10--20\% level in portions of our parameter space.
However, as the cooling rates of the dominant coolants often have 
uncertainties that are comparable to or larger than the size of the 
contributions from these minor coolants, the accuracy we would gain by
including them is less than might be expected, and does not (in our opinion)
justify the additional complexity and computational expense that would be
required in order to treat them.

Finally, it is clear that our conclusions here are sensitive to our choice of
elemental abundance ratios. For instance, an increase in the iron abundance
(relative to the other metals) of 1--2 orders of magnitude would render 
Fe and ${\rm Fe^{+}}$ important coolants in portions of our parameter 
space, necessitating their inclusion in our model. Such an increase 
would be expected if the enrichment of the gas were dominated by 
pair-instability supernovae with masses close to the top end of the 
140--$260 \: {\rm M_{\odot}}$ allowed mass range \citep{hw02}. 
However, studies of the abundance ratios found in extremely 
metal-poor stars in the Galactic halo, which at present give us the
best picture available of the elemental composition of very metal-poor
gas, do not find evidence for significant enrichment by pair-instability
supernovae \citep[see e.g.\ the discussion in][]{tvs04} and so at the present 
time there is no compelling reason to include iron in our chemical model.

Our chemical model therefore consists of the eight coolants discussed above
-- $\mH$, $\mHt$, $\hd$, $\mC$, $\cp$, $\mO$, $\mSi$ and $\sip$ -- together
with ten additional species that play key roles in determining one or more of 
the abundances of the coolants: $\me$, $\Hp$,  $\Hm$, $\mHtp$, $\He$, 
$\Hep$, $\mD$, $\Dp$, $\op$ and $\sipp$. Our rationale for including $\He$
and $\Hep$ in our model is that in the presence of a significant flux of 
hard UV photons, X-rays or cosmic rays, ionized helium can act as an 
important source of free electrons, and moreover can transfer charge
to neutral carbon or silicon (but not oxygen) far more effectively than 
$\Hp$ can.

We do not include minor primordial coolants such as ${\rm LiH}$ or ${\rm H_{3}^{+}}$.
These are never important at low densities and rarely important at high densities
(see \citealt{mon05}, \citealt{GS06,GS07}). More significantly, we do not 
include molecular coolants such as CO or ${\rm H_{2}O}$, assuming instead
that the bulk of the carbon and oxygen in the gas remains in atomic or ionized form.
This assumption dramatically simplifies the chemical modelling of the gas, but at
the same time restricts the range of physical conditions over which the resulting 
model is useful. 

\begin{figure}[Htb]
\centering
\epsfig{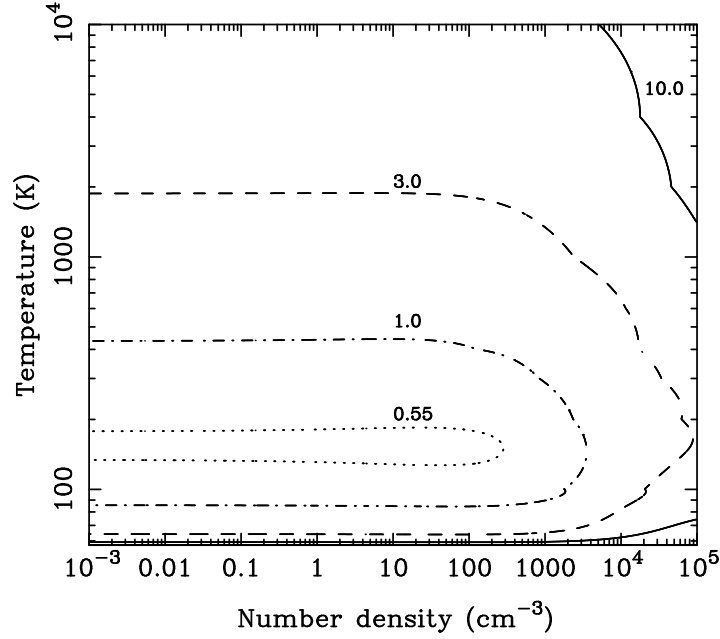} 
\caption{Value of $f_{\rm mol}$ (see Equation~\ref{fmol_eq}) as a function of 
temperature and density. Contours corresponding to $f_{\rm mol} = 0.55$
({\it dotted}), 1.0 ({\it dot-dashed}), 3.0 ({\it dashed}) and 10.0 ({\it solid}) are
plotted. We see that $f_{\rm mol} < 10$ for almost the whole of the parameter 
space considered. \label{fmol}}
\end{figure}

To assess the conditions for which this approximation is justified, we need to
know two things. First, what fraction of the available carbon and oxygen must
be locked up in CO or ${\rm H_{2}O}$ in order for cooling from these 
molecules to dominate over fine structure cooling? Second, under what 
conditions are these fractions achievable within a dynamically interesting 
timescale? To answer the first of these questions, we have performed 
calculations using the treatment of carbon and oxygen fine structure cooling
discussed in \S\ref{fscool} below, together with a table-based treatment of CO and 
${\rm H_{2}O}$ taken from \citet{nk93} and \citet{nlm95}. We have computed
$f_{\rm mol}$, defined as
\begin{equation}
f_{\rm mol} = \frac{\Lambda_{\rm CO} + \Lambda_{\rm H_{2}O}}{\Lambda_{\cp} 
+ \Lambda_{\mC} + \Lambda_{\mO}},  \label{fmol_eq}
\end{equation}
where $\Lambda_{i}$ is the cooling rate per unit volume due to species $i$,
for a wide range of temperatures and densities. We assume that 
$n_{\mH} \gg \rm{max}(n_{\mHt}, n_{\rm e})$ and that all of the cooling occurs
in the optically thin regime. We adopt a nominal redshift $z=20$ and consider
only temperatures $T > T_{\rm CMB}(z) = 57.2 \: {\rm K}$, under the 
assumption that heating from the CMB will prevent the gas from cooling
appreciably below this temperature. In Figure~\ref{fmol}, we show how 
$f_{\rm mol}$ varies as a function of temperature and density. 

Figure~\ref{fmol} demonstrates that over most of the parameter space that we 
have examined, $f_{\rm mol}$ is of order unity. We find that $f_{\rm mol} > 10$
only for temperatures very close to the CMB temperature (which is likely an
numerical artifact, a result of the fact that our treatment 
of fine structure cooling includes the effects of radiative pumping by the CMB, 
while our treatment of CO and ${\rm H_{2}O}$ cooling does not), and at high 
densities, where the level populations of the fine structure coolants start to 
reach their local thermodynamic equilibrium values, causing the fine structure 
cooling rate per atom to saturate. For CO or ${\rm H_{2}O}$
cooling to be effective, we therefore require that about as many carbon and
oxygen atoms be incorporated into molecules as remain in atomic form. 

With these values of $f_{\rm mol}$ in hand, we can now turn to the question of 
whether it is possible to produce enough CO and ${\rm H_{2}O}$ in the gas
within an interesting timescale. Since we require a significant amount of oxygen
to be in the form of coolant molecules, the timescale of interest, $t_{\rm conv}$,
is given approximately by
\begin{equation} 
 t_{\rm conv} \simeq \frac{n_{\mO}}{R_{\rm f}}
\end{equation}
where $R_{\rm f}$ is the net rate of formation of coolants per unit volume.
Therefore, to estimate $t_{\rm conv}$ we must first estimate $R_{\rm f}$.

Although a complete discussion of the formation and destruction mechanisms 
of CO and ${\rm H_{2}O}$ in low metallicity gas is beyond the scope of this paper, 
we will briefly summarize the most important points. In hot gas 
($T \simgreat 600 \: {\rm K}$), most CO and water molecules form via reaction pathways
initiated by the hydroxyl radical, OH \citep[see e.g.][]{HOL79,WAG87}.
This is formed by
\begin{equation}
\mO + \mHt \rightarrow {\rm OH} + \mH, \label{ohf1}
\end{equation}
but most is then destroyed by
\begin{equation}
\oh + \mH \rightarrow \mO + \mHt.  \label{ohd1}
\end{equation}
However, a small fraction instead reacts to form other molecular species,
such as water or CO, e.g.\
\begin{eqnarray}
\oh + \mHt & \rightarrow & {\rm H_{2}O} + \mH, \label{ohd2} \\  
\oh + \mC & \rightarrow & {\rm CO} + \mH. \label{ohd3}
\end{eqnarray}
The resulting ${\rm H_{2}O}$ and CO molecules can be destroyed by collisions
with atomic hydrogen:
\begin{eqnarray}
{\rm H_{2}O} + \mH & \rightarrow & \oh + \mHt, \label{water1} \\
{\rm CO} + \mH & \rightarrow & \mC + \oh. \label{cod1}
\end{eqnarray}
If, as will generally be the case in low-metallicity gas, $x_{\mHt} \simless 0.1$
(where $x_{\mHt}$ is the fractional abundance of $\mHt$ relative to the total number 
of hydrogen nuclei), then the destruction of water by reaction~\ref{water1} is far more 
effective than its formation by reaction~\ref{ohd2}, and water will never account
for more than a small fraction of the available oxygen. On the other hand,
the CO formed by reaction~\ref{ohd3} can potentially account for almost all of the
available oxygen or carbon -- whichever is present in the smaller amount -- as
the destruction of CO by reaction~\ref{cod1} is ineffective at $T < 5000 \: {\rm K}$.
Therefore, the net rate of formation of coolant molecules (primarily ${\rm CO}$) 
is given approximately by
\begin{equation}
R_{\rm f} \simeq \frac{ k_{\ref{ohf1}} k_{\ref{ohd3}} x_{\mC}}{k_{\ref{ohd1}} 
x_{\mH} + k_{\ref{ohd3}} x_{\mC}} n_{\mO} n_{\mHt}  \: {\rm cm^{-3}} \: {\rm s^{-1}},
\end{equation}
where $k_{i}$ is the rate coefficient of reaction $i$, $x_{\mH}$ and $x_{\mC}$
are the fractional abundances of atomic hydrogen and atomic carbon, and 
$n_{\mO}$ and $n_{\mHt}$ are the number densities of $\mO$ and $\mHt$ 
respectively. The timescale to convert significant quantities of oxygen to CO 
is then given approximately by
\begin{eqnarray}
t_{\rm conv} & \simeq & \frac{n_{\mO}}{R_{\rm f}}, \\
 & = & \frac{1}{n_{\mHt}} \frac{k_{\ref{ohd1}} x_{\mH} + k_{\ref{ohd3}} x_{\mC}} 
{ k_{\ref{ohf1}} k_{\ref{ohd3}} x_{\mC}}.
\end{eqnarray}
For $x_{\mHt} = 10^{-3}$, which is a reasonable value for low metallicity gas if
$\mHt$ formation on dust is unimportant, this gives a timescale of approximately
\begin{equation}
t_{\rm conv} \simeq \frac{100}{n} \frac{{\rm Z_{\odot}}}{{\rm Z}} \: {\rm Myr},
\end{equation}
where $n$ is the number density of hydrogen nuclei, and where we have 
adopted the values for the various rate coefficients that are given in \citet{TEU00}.

In cold gas, all of these reactions (except for reaction~\ref{ohd3}) are ineffective,
and other processes dominate the formation of water and CO. A good summary
of the relevant chemistry is given in \citet{BLA77}. When $x_{\op} x_{\mHt} 
\simgreat 10^{-9} x_{\mO} x_{\mH}$, 
the most important mechanism involves the formation of the $\oh^{+}$ ion via the rapid 
ion-neutral reaction
\begin{equation}
\op + \mHt \rightarrow \oh^{+} + \mH. \label{ohpf}
\end{equation}
If $x_{\rm e} > 1.6 \times 10^{-3} T^{1/2} x_{\mHt}$, then most of the resulting $\oh^{+}$
ions simply dissociatively recombine:
\begin{equation}
\oh^{+} + \me \rightarrow \mO + \mH. \label{ohpdr}
\end{equation}
Otherwise, they can then react further with $\mHt$ to give 
${\rm H_{2}O^{+}}$ and ${\rm H_{3}O^{+}}$, with dissociative recombination
of the latter producing $\oh$ and water. CO formation follows through 
reaction~\ref{ohd3}. All of these reactions occur rapidly, and so  the
net rate of formation of coolant molecules is given by the rate of formation of
$\oh^{+}$ multiplied by the fraction of $\oh^{+}$ that is not destroyed by 
dissociative recombination, (1 - $f_{\rm dr})$:
\begin{equation}
 R_{\rm  f} \simeq k_{\ref{ohpf}} (1.0 - f_{\rm dr}) n_{\op} n_{\mHt},
\end{equation}
For $x_{\rm e} = 1.6 \times 10^{-3} T^{1/2} x_{\mHt}$, we have $f_{\rm dr} = 0.5$,
and hence $R_{\rm f} = 0.5 k_{\ref{ohpf}} n_{\op} n_{\mHt}$. In that case,
the time required to convert most of the oxygen to $\oh^{+}$ and thence to other 
molecules is
\begin{equation} 
t_{\rm conv} \simeq \frac{1}{k_{\ref{ohpf}} x_{\mHt} x_{\Hp} n},
\end{equation}
where we have used the fact that $n_{\op}/n_{\mO} \simeq n_{\Hp}/n_{\mH}$
owing to the rapid transfer of charge between oxygen and hydrogen 
(see \S\ref{metal_chem} below). If we assume that $x_{\rm e} \simeq x_{\Hp}$
and that $x_{\mHt} = 10^{-3}$, then this gives us a value for $t_{\rm conv}$ of
\begin{equation}
t_{\rm conv} \simeq \frac{1000}{n} \: {\rm Myr}.
\end{equation}
Decreasing the fractional ionization of the gas will increase $t_{\rm conv}$,
but increasing it will have no significant effect.

Finally, if the fractional ionization of the gas is too low for reaction~\ref{ohpf}
to operate effectively, then the formation of coolant molecules again occurs
primarily via hydroxyl, which in this case is formed mainly by direct radiative
association \citep{JUL71,SMI76}
\begin{equation}
\mO + \mH \rightarrow \oh + \gamma. \label{ohra}
\end{equation}
In this case, $t_{\rm conv}$ is simply
\begin{eqnarray}
t_{\rm conv} & = & (k_{\ref{ohra}} n)^{-1}, \nonumber \\
 & \simeq & \frac{40}{n} \: {\rm Gyr}. 
\end{eqnarray}

Comparing the three values of $t_{\rm conv}$ derived above, we see that
in gas with ${\rm Z} = 0.1 \: {\rm Z_{\odot}}$, $t_{\rm conv} \simeq 1000
n^{-1} \: {\rm Myr}$, regardless of the gas temperature, provided that 
$x_{\mHt} \simeq 10^{-3}$ and that the fractional ionization satisfies
the constraint given above. If we compare this with the characteristic
physical timescale of the problem of interest, $t_{\rm char}$, then it is
simple to show that for densities
\begin{equation}
n \simless \frac{1000 \: {\rm Myr}}{t_{\rm char}} \: {\rm cm^{-3}},
\end{equation}
CO and ${\rm H_{2}O}$ will not form in quantities large enough to
dominate the cooling, and hence it is valid to ignore these molecules 
and all of their associated chemistry. Note also that if we were to include 
the effects of photodissociation of $\oh$, ${\rm H_{2}O}$ and CO in the 
above analysis, then this would push the required density to an even 
larger value.

In a gravitationally collapsing protogalaxy, a reasonable value for 
$t_{\rm char}$ is the gravitational free-fall timescale $t_{\rm ff}$, and
in this scenario, our neglect of the molecular chemistry is valid as
long as $n < 400 \: {\rm cm^{-3}}$. On the other hand, if we want
to simulate the thermal evolution of the interstellar medium in a metal-poor dwarf
galaxy, a more reasonable timescale may be the sound-crossing
time of the disk, which is of order $100 \: {\rm Myr}$ for a 
$1 \: {\rm kpc}$ disk and a sound speed of $10 \: {\rm km} \: 
{\rm s^{-1}}$. In this case, our model is valid only for 
$n < 10 \: {\rm cm^{-3}}$.

As a consistency check on our conclusions here, we examined the results
of \citet{OMU05}, who model the thermal and chemical evolution of 
freely-falling gas at a range of metallicities far below solar, using a detailed
treatment of the gas chemistry. They find that for $Z = 0.01 \: {\rm Z_{\odot}}$,
significant conversion of carbon and oxygen to molecular form does not
occur until $n > 10^{3} \: {\rm cm^{-3}}$, in line with the value derived here.
At lower metallicities, an even higher gas density is required.  As another
check, we have computed the evolution of the  ${\rm CO}$ and ${\rm H_{2}O}$ 
abundances in the gas at the center of several of the simulated 
protogalactic halos discussed in \citeauthor{jgkm07}~(2007; hereafter paper II),
using values for the density, temperature, $\mHt$ abundance and $\Hp$ abundance 
taken from our simulations,  and modelling the chemistry with the full UMIST99 chemical 
network \citep{TEU00}. We find that in most of these runs, our neglect of 
the molecular coolants is justified, as their abundances never become large
enough for them to significantly affect the cooling. Our approximation begins
to break down in the runs with ${\rm Z} = 0.1 \: {\rm Z_{\odot}}$, where about 
10\%--20\% of the total carbon and oxygen are incorporated into CO and 
${\rm H_{2}O}$, which is just enough to affect the cooling at high temperatures 
and/or high densities. However, only in our runs with ${\rm Z} = {\rm Z_{\odot}}$ 
does it break down completely. Therefore,  the use of our highly simplified 
chemical model would appear to be justified in low-density gas with 
${\rm Z} \leq 0.1 \: {\rm Z_{\odot}}$.

We are thus left with a set of eighteen chemical species that must be modelled:
$\me$, $\Hp$, $\mH$, $\Hm$, $\mHtp$, $\mHt$, $\Dp$, $\mD$, $\hd$, $\He$,
$\Hep$, $\mC$, $\cp$, $\mO$, $\op$, $\mSi$, $\sip$ and $\sipp$. 
The combined evolution of the abundances of these 
species is described by a chemical network consisting of 74 reactions: 
47 collisional gas-phase reactions (summarized in Table~\ref{tab:chem_gas_coll}),
12 photochemical gas-phase reactions (summarized in Table~\ref{tab:chem_gas_photo}), 
7 grain surface reactions (summarized in Table~\ref{tab:chem_grain}), and
8 reactions involving cosmic rays (summarized in Table~\ref{tab:cosmic}). The 
abundances of these species are also constrained by seven conservation laws:
\begin{eqnarray}
x_{\Hp} + x_{\mHtp} + x_{\Dp} + x_{\Hep} + x_{\cp} + x_{\op} + x_{\sip}  + x_{\sipp}&  
= &  x_{\rm e} + x_{\Hm},  \label{chg_cons} \\ 
x_{\Hp} + x_{\mH} + x_{\Hm} + 2 x_{\mHtp} + 2 x_{\mHt} + x_{\hd} & = &  1, \\
x_{\Dp} + x_{\mD} + x_{\hd} & = &  x_{\rm D, \, tot}, \\
x_{\Hep} + x_{\He} & = & x_{\rm He, \, tot}, \\
x_{\mC} + x_{\cp} & =  & x_{\rm C, \, tot}, \\
x_{\mO} + x_{\op} & =  & x_{\rm O, \, tot}, \\
x_{\mSi} + x_{\sip} + 2 x_{\sipp} & =  & x_{\rm Si, \, tot} \label{si_cons}
\end{eqnarray}
where $x_{i}$ is the fractional abundance of chemical species $i$ relative to the total
abundance of hydrogen nuclei, and where $ x_{\rm D, \, tot}$, $x_{\rm He, \, tot}$, 
$x_{\rm C, \, tot}$,  $x_{\rm O, \, tot}$, and $x_{\rm Si, \, tot}$ are the total abundances of 
deuterium, helium, carbon, oxygen and silicon, respectively. Furthermore, we assume 
in our modelling that $\Hm$ and $\mHtp$ are in chemical equilibrium, allowing us to write 
their abundances as: 
\begin{equation}
x_{\Hm} = \frac{k_{1} x_{\mH} x_{\rm e} n}{(k_{2} x_{\mH} + k_{5} x_{\Hp} + k_{15} x_{\rm e} 
+ k_{16} x_{\mH} + k_{17} x_{\Hp}) n + R_{51}},
\end{equation}
and
\begin{equation}
x_{\mHtp} = \frac{(k_{3} x_{\mH} x_{\Hp} + k_{7} x_{\mHt} x_{\Hp} + k_{17} x_{\Hm} x_{\Hp})n + \zeta_{\mHt} x_{\mHt} 
+ R_{54} x_{\mHt}}{(k_{4} x_{\mH} + k_{6} x_{\rm e}) n + R_{52}}, \label{eq_h2p}
\end{equation}
where $n$ is the number density of hydrogen nuclei. This assumption is generally
justified in simulations of the cooling and gravitational collapse of gas in protogalactic
halos, as the timescales on which $\Hm$ and $\mHtp$ reach chemical equilibrium 
are much shorter than the cooling or free-fall timescale. (For a more detailed discussion 
of this point, see \citealt{GLO06}).

The constraints represented by equations~\ref{chg_cons}--\ref{eq_h2p} allow us to
reduce the total number of chemical rate equations that must be solved to only nine. 
In practice, we generally choose to solve for the abundances of the ionized species
($\Hp$, $\Dp$, $\Hep$, $\cp$, $\op$, $\sip$, $\sipp$), $\mHt$ and $\hd$, but alternative 
choices are possible and would not significantly alter the results obtained.

\subsection{Selection of reactions}
\label{react_select}
The number of chemical reactions that could be included in our chemical network is 
very large, despite the limited number of chemical species involved. Fortunately, 
many of these reactions have little or no impact on the evolution of the abundances 
of our main coolants and so the number of reactions that need to be included in our
chemical network remains reasonably small. 

We can divide the reactions that must be included into two subsets. The first 
subset consists of the reactions required to model the chemistry of hydrogen,
helium and deuterium, including the formation and destruction of $\mHt$ 
and $\hd$ (reactions 1--29, 48--55, 60--63, 67--70). The second subset consists of 
the reactions required to model the carbon, oxygen and silicon chemistry 
(reactions 30--47, 56--59, 64--66, 71--74).

\subsubsection{Hydrogen, helium and deuterium chemistry}
The amount of $\Hp$ present in the gas is controlled by seven main reactions:
collisional ionization of $\mH$ by electrons (reaction 11), charge transfer
with helium (reactions 26 \& 27), photoionization (reaction 48), cosmic ray
ionization (reaction 67), gas-phase
recombination (reaction 13) and recombination on the surface of dust 
grains (reaction 61). Similar reactions (nos.\ 12, 28, 29, 49, 68, 14 \&  62) partially 
determine the $\Dp$ abundance, but in this case, charge transfer to and from 
hydrogen (reactions 18 \& 19) is also of great importance, owing to the very 
large abundance of hydrogen relative to deuterium. Finally, the $\Hep$
abundance is controlled primarily by collisional ionization (reaction 24),
photoionization (reaction 50), cosmic ray ionization (reaction 69),
gas-phase and grain-surface recombination (reactions 25 \& 63) and charge 
transfer with hydrogen (reactions 26 \& 27).

The remaining 24 reactions in this subset control the formation and 
destruction of $\mHt$ and $\hd$. $\mHt$ forms in the gas phase via the 
intermediate ions $\Hm$ and $\mHtp$ (reactions 2 \& 4), as well as on the 
surface of dust grains (reaction 60). It is destroyed by collisions with $\Hp$, 
$\me$, $\mH$ and $\mHt$ (reactions 7--10), and can also be photodissociated
or photoionized by UV radiation (reactions 53 \& 54), or ionized by cosmic rays
(reaction  70). Collisions with 
$\He$  \citep{drcm87}
\begin{equation}
\mHt + \He \rightarrow \mH + \mH + \He,
\end{equation}
and $\Hep$ \citep{ba84}
\begin{eqnarray}
\mHt  + \Hep & \rightarrow & \mH + \Hp + \He, \\
 & \rightarrow & \mHtp + \He,
\end{eqnarray}
can also destroy $\mHt$, but in general these processes are not as effective
as collisions with hydrogen, and so they can be omitted from our simplified 
chemical model without significantly affecting its accuracy.

The $\Hm$ and $\mHtp$ ions required for gas-phase $\mHt$ formation are 
formed primarily by the radiative association of atomic hydrogen with
free electrons or protons respectively (reactions 1 \& 3), and while
reactions 2 \& 4 generally dominate the removal of $\Hm$ and $\mHtp$
from the gas, in hot or highly ionized gas a number of other processes
become competitive (reactions 5, 6 \& 15-17) Photodissocation of $\Hm$ 
and $\mHtp$ (reactions 51 \& 52) can also become important if the incident 
radiation field is strong. 

Finally, although $\hd$ can form from intermediate ions such as ${\rm D}^{-}$ or
${\rm HD^{+}}$ in a manner analogous to $\mHt$ \citep[see e.g.][]{sld98}, most 
actually forms from $\mHt$ via reaction 20:
\begin{equation}
\mHt + \Dp \rightarrow \hd + \Hp.
\end{equation}
The resulting $\hd$ can be destroyed by the inverse of this reaction 
(reaction 21), or by photodissociation (reaction 55). In hot gas, the
$\hd$ abundance is also influenced by reactions between $\mHt$ and 
$\mD$ and $\hd$ and $\mH$ (nos.\  22 \& 23 respectively). Note that with the 
exception of the grain surface reactions (which are discussed in 
\S\ref{surface_chem} below), the reactions required to model the 
hydrogen, helium and deuterium chemistry accurately in metal-enriched 
gas are just the same as those required to model primordial gas. 

The values for many of these rate coefficients are known to within a small
amount of uncertainty at the temperatures and densities of interest, and so
our choice of the particular values used here should be uncontroversial. 
However, a few of our assumptions demand further comment. 

First, we note that the rates listed for reactions involving $\mHt$ or $\hd$
as a reactant generally assume that these molecules are not vibrationally
excited. This assumption is reasonable at the low densities treated here,
but breaks down at densities $n \simgreat 10^{4} \: {\rm cm^{-3}}$.

Second, we note that for several reactions involving deuterium where no 
readily available rate coefficient exists in the astrophysical literature, we 
have assumed that  the rate is the same as for the analogous reaction 
involving $\mH$ or $\Hp$. 

The rates of processes involving cosmic rays (see Table~\ref{tab:cosmic}) 
depend on the energy spectrum and energy density of cosmic rays. 
These are poorly known at the relevant energies even in the local ISM,
and far less is known concerning high-redshift cosmic rays. For this
reason, we do not give absolute values for these rates, but instead
parameterize them in terms of the cosmic ray ionization rate for atomic
hydrogen, $\zeta_{\rm H}$, which can then be considered an adjustable
parameter of the model. 

Finally, we note that two of the reactions that regulate the $\Hm$ abundance
have large uncertainties in their rate coefficients. The reactions in question
are the associative detachment of $\Hm$ with $\mH$ (reaction 2) and the
mutual neutralization of $\Hm$ with $\Hp$ (reaction 5). As discussed in
\citet{GLO06}, the uncertainties in the rates of both of these reactions may 
be as large as an order of magnitude. In gas with a high fractional
ionization, this uncertainty can lead to a significant uncertainty in the $\mHt$
formation rate and in the final $\mHt$ fractional abundance, particularly
in the presence of a strong ultraviolet background radiation field. In the
chemical model presented here, we have followed \citet{GP98} and have 
adopted a rate coefficient for reaction 2 taken from \citet{LAU91} and a rate 
coefficient for reaction 5 taken from \citet{MOS70}. However, we caution the 
reader that this should not be regarded as an endorsement of the accuracy
of these particular values.

\subsubsection{Carbon, oxygen and silicon chemistry}
\label{metal_chem}
As discussed in \S\ref{chem_choice} above, we do not include the molecular
chemistry of these elements in this simplified model. Our treatment of the carbon, 
oxygen and silicon chemistry is therefore purely a treatment of the charge balance 
of these species. 

We begin with oxygen, in many respects the simplest of the three to treat. The
ionization potential of neutral oxygen is only 0.02~eV larger than that of neutral 
hydrogen, and so charge transfer between $\op$ and $\mH$ or $\Hp$ and $\mO$ 
(reactions 36 \& 37) occurs rapidly. Since the hydrogen abundance is orders of 
magnitude larger than the oxygen abundance, this means that the ratio of ionized to 
neutral oxygen is controlled by the ratio of ionized to neutral hydrogen,
i.e.\ that 
\begin{equation}
\frac{x_{\op}}{x_{\mO}} \simeq \frac{k_{37}}{k_{36}}  \frac{x_{\Hp}}{x_{\mH}}. 
\end{equation}
In most circumstances, reactions 36 \& 37 are the only reactions required in order 
to accurately model the oxygen chemistry. Nevertheless, for completeness we
also include several other processes: radiative recombination (reaction 32),
grain surface recombination (reaction 65), collisional ionization (reaction 35), 
charge transfer with $\Hep$ (reaction 38), photoionization (reaction 57) and
cosmic ray ionization (reaction 72).

In the case of carbon, the situation is rather different. First, charge transfer
from $\Hp$ to $\mC$ (reaction 39) is much less effective than charge transfer from 
$\Hp$ to $\mO$, and so reaction 39 plays a far less important role in the carbon
chemistry than reaction 37 does in the oxygen chemistry. Second, carbon has an
ionization potential of only 11.26~eV. This means that charge transfer from $\cp$
to $\mH$ is significantly endothermic, rendering it unimportant at $T < 10^{4} \: {\rm K}$
(although we include it here for completeness). It also means that neutral carbon 
can be photoionized by ultraviolet photons with wavelengths $\lambda > 912$~\AA, 
which can penetrate easily into low metallicity protogalactic gas. Consequently, 
photoionization of $\mC$ (reaction 56) plays an important role in the carbon chemistry, 
whereas photoionization of $\op$ is unimportant outside of $\hii$ regions. In addition to 
charge transfer and photoionization, ionized carbon can also be produced by collisional 
ionization (reaction 33), although this is important only for temperatures 
$T > 9000 \: {\rm K}$, by charge transfer with $\Hep$ (reaction 41) and by cosmic ray
ionization (reaction 71).  $\cp$ is removed 
from the gas primarily by recombination in the gas phase (reaction 30) and on grain 
surfaces (reaction 64), although if the free electron abundance is small, charge transfer 
from $\cp$ to $\mSi$ (reaction 44) can also become important.

In the case of silicon, a similar set of processes operate: collisional ionization (reaction
34), cosmic ray ionization (reaction 73), gas-phase recombination (reaction 31), 
grain surface recombination (reaction 66),
charge transfer from $\Hp$, $\Hep$ or $\cp$ to $\mSi$ (reactions 42, 43 \& 44), and
photoionization by an ultraviolet radiation field (reaction 58), assuming one is present.
In the case of silicon, however, we also include the doubly ionized ion, $\sipp$, in our 
chemical model, since this can be produced by charge transfer from $\Hp$ to $\sip$
(reaction 45) with an endothermicity of only a few eV, rendering it potentially important 
at temperatures $\sim 10^{4} \: {\rm K}$. It can also be produced by photoionization
(reaction 59), although in this case the 16.35~eV energy requirement renders this process
important only within \hii regions, or by cosmic ray ionization (reaction 74).
The $\sipp$ produced by these processes can be destroyed by charge transfer with 
neutral hydrogen (reaction 46), or  by recombination (reaction 47). 

Regarding the accuracy of our adopted reaction rates, we note that while
many have well determined rate coefficients, there are two notable exceptions.
The only available rate coefficients for reactions 43 and 44, which involve the 
transfer of charge from $\Hep$ to $\mSi$ and from  $\cp$ to $\mSi$ respectively,
assume that the reactions both  proceed at the Langevin
rate. In practice, highly exothermic charge transfer reactions often
proceed at a rate far below the simple Langevin rate, and so the rates
of these reactions may be overestimated in our model.

\subsection{Grain surface chemistry}
\label{surface_chem}
If dust grains are present in the metal-enriched gas, then our chemical model must
account for the effects of reactions occurring on the surface of the grains, as these
reactions are often far more effective than their gas-phase counterparts. We therefore
include a small number of grain surface reactions in our model, as summarized in 
Table~\ref{tab:chem_grain}. 

The only grain surface reaction between neutral species that is included in our model
is the formation of molecular hydrogen (reaction 60). This is a hugely important reaction
in the local interstellar medium, since at typical interstellar densities it is the only process 
capable of producing the large quantities of molecular hydrogen that are directly observed 
\citep[see e.g.][]{WAK06} or inferred from other molecular tracers, such as CO. Theoretical 
modeling \citep{HIR02,GLO03,CAZ04} suggests that it remains important down to metallicities 
of order $10^{-3} \: {\rm Z_{\odot}}$, within the range of applicability for the chemical model
presented here.

The rate that we adopt for this process is based on the widely used rate of \citet{HOL79}.
This was derived for solar metallicity gas, assuming a distribution of grain sizes as given
in \citet{MAT77}, and to adapt it for use in low metallicity gas, we simply assume that
the rate scales linearly with the metallicity ${\rm Z}$. The validity of this assumption is
open to question, as many of the features of the grain population, such as the grain size 
distribution or the mix of compositions, may differ greatly between Milky Way dust and
protogalactic dust. An alternative, physically motivated approach would be to adopt a 
grain size distribution and mix of compositions based on the results of numerical modeling 
of dust formation in high redshift supernovae \citep{TOD01,NOZ03,SCH04} 
and then to compute the $\mHt$ formation rate expected for this grain population. This is 
the approach used by \citet{SCH06}.
However, the uncertainties associated with this approach are considerable. To begin
with, the predictions of the numerical models are highly sensitive to the degree of 
mixing assumed to occur within the supernova ejecta \citep{NOZ03}. In addition, dust 
destruction in the reverse shock is typically not taken into account in  these models, 
and to the best of our knowledge, the amount of dust that survives the passage of the 
shock has yet to be fully quantified (although research in this area is actively
proceeding; R. Schneider, priv.\ comm.). Finally, the changes wrought on the grain 
population by subsequent processing in the interstellar medium or intergalactic medium 
\citep[see e.g.][]{VEN06} are not  fully understood. In light of these uncertainties, 
we do not believe that this approach currently offers much of an advantage over our 
simple assumption of an $\mHt$ formation rate that scales with metallicity. 

We do not include any neutral-neutral surface reactions in our model other than
$\mHt$ formation. For neutral-neutral surface reactions to be able to significantly
affect the ability of the gas to cool, they must be able to alter the abundances of 
neutral carbon, oxygen or silicon by a substantial amount. We can place an upper
limit on the rate at which these reactions occur if we assume that the reaction 
probability $f_{\rm react} = 1$, i.e.\ that every collision between a metal atom and 
a grain results in a reaction. In that case, the reaction rate per unit volume for an 
atomic species $i$ is given by 
\begin{equation}
 R_{\rm i} = v_{{\rm th}, i} A  n_{\rm i},  \label{grain_rate}
\end{equation}
where $v_{{\rm th}, i}$ is the thermal velocity of atomic species $i$, and $A$ is the 
total surface area of grains per unit volume of gas. For Milky Way dust of the 
type assumed by \citet{HOL79}, $A \simeq 3 \times 10^{-21} n \: {\rm cm^{-1}}$,
where $n$ is the number density of hydrogen nuclei. At lower metallicity, our
assumption that the grain size distribution does not change with metallicity
implies that the value of $A$ in gas with a metallicity $Z$ is simply 
$A \simeq 3 \times 10^{-21} ({\rm Z} / {\rm Z_{\odot}}) n \: {\rm cm^{-1}}$.
We can therefore rewrite equation~\ref{grain_rate} as
\begin{equation}
 R_{\rm i} \simeq 4.7 \times 10^{-17} \left(\frac{T}{m_{i}}\right)^{1/2} 
 \left(\frac{\rm Z}{\rm Z_{\odot}}\right) n \, n_{\rm i}, \label{gr2}
 \end{equation}
where $m_{i}$ is the mass of species $i$ in atomic mass units. The 
corresponding conversion timescale $t_{\rm conv} = n_{\rm i} / R_{\rm i}$ is then
\begin{equation}
t_{\rm conv} = 2.1 \times 10^{16} n^{-1} \left(\frac{T}{m_{i}}\right)^{-1/2} 
 \left(\frac{\rm Z}{\rm Z_{\odot}}\right)^{-1} \: {\rm s}.
\end{equation}
If $f_{\rm react} < 1$, then this expression becomes:
\begin{equation}
t_{\rm conv} = 2.1 \times 10^{16} f_{\rm react}^{-1} n^{-1} \left(\frac{T}{m_{i}}\right)^{-1/2} 
 \left(\frac{\rm Z}{\rm Z_{\odot}}\right)^{-1} \: {\rm s}.
\end{equation}
If we equate this to a characteristic physical timescale $t_{\rm char}$, then we
can show that $t_{\rm conv} > t_{\rm char}$ as long as
\begin{equation}
n < 665 \left(\frac{1 \: {\rm Myr}}{t_{\rm char}} \right) f_{\rm react}^{-1} 
\left(\frac{T}{m_{i}}\right)^{-1/2}  \left(\frac{\rm Z}{\rm Z_{\odot}}\right)^{-1}
\: {\rm cm^{-3}}.
\end{equation}
For gas with a temperature $T = 2000 \: {\rm K}$ and metallicity 
${\rm Z} = 0.1 \: {\rm Z_{\odot}}$, and with atomic carbon as the 
colliding species, this gives:
\begin{equation}
n < \frac{500 f_{\rm react}^{-1} \: {\rm Myr}}{t_{\rm char}} \: {\rm cm^{-3}}.
\end{equation}
At higher temperatures, the limiting $n$ will be slightly smaller, but
at $T > 2000 \: {\rm K}$, collisional dissociation of most molecular
species is highly effective, and so in this temperature regime, grain
surface reactions are unlikely to be important.

From this analysis, we see that even if $f_{\rm react} = 1$, neutral-neutral grain 
surface reactions are unimportant at gas densities $n < (500 \: {\rm Myr} / 
t_{\rm char}) ({\rm Z} / 0.1 \: {\rm Z_{\odot}})^{-1} \: {\rm cm^{-3}}$. In gravitationally 
collapsing gas with $t_{\rm char} = t_{\rm ff}$, this corresponds to 
$n \simless 100 ({\rm Z} / 0.1 \: {\rm Z_{\odot}})^{-1} \: {\rm cm^{-3}}$. Note that a key 
point here is that for neutral-neutral grain surface reactions involving our atomic 
coolants (C, O etc.) to significantly affect the thermal behaviour of the gas, a large 
fraction of the total number of coolant atoms must react, whereas for grain surface 
reactions to affect the $\mHt$ cooling rate by significantly altering the $\mHt$ 
abundance, only a small fraction of the total number of hydrogen atoms must 
react. This means that in the case of $\mHt$, the relevant conversion timescale 
is several orders of magnitude shorter, and hence our density limit would be
significantly smaller if we were to omit grain surface $\mHt$ formation from our
model.

One way in which this simple analysis could break down is if $f_{\rm react}$ were
much larger for reactions involving the coolant atoms than for $\mHt$ formation.
However, a recent analysis of $\mHt$ formation on grain surfaces by \citet{ct04}
that takes both physisorbed (i.e.\ van der Waals bonded) and chemisorbed (i.e.\
chemically bonded) hydrogen into account demonstrates that in the conditions
of interest in this paper, $f_{\rm react} \sim 1$ (with the result that the computed 
$\mHt$ formation rate is very similar to the widely-used rate of \citet{HOL79} that 
is used in our model). A comparable analysis has not been performed for 
reactions involving C, O or Si, but clearly they cannot have $f_{\rm react} > 1$,
and so there is little scope for these reactions to occur significantly faster than
$\mHt$ formation.

Finally, we include in our surface chemistry model six important reactions involving
ions: the recombination of $\Hp$, $\Dp$, $\Hep$, $\cp$, $\op$ and $\sip$ with electrons 
on the surface of grains. As grain surface recombination is a non-radiative process, it 
proceeds at a much faster rate than radiative recombination in the gas phase. Moreover, 
if a typical grain is negatively charged, then the effective cross-section for collisions will 
be much enhanced over the geometric cross-section due to Coloumb focussing. Grain 
charging is largely determined by the parameter \citep{BAK94}
\begin{equation}
\psi = \frac{G \sqrt{T}}{n_{\rm e}},
\end{equation}
where $G \simeq 0.01 J_{21}$ is a measure of the radiation energy density between 
$6 \: {\rm eV}$  and $13.6 \: {\rm eV}$ relative to the \citet{habing68} field. When 
$\psi$ is small, most grains are negatively charged,  and so in these conditions grain 
surface recombination can be important even if neutral-neutral grain surface reactions 
are unimportant. 

To model grain surface recombination, we follow \citet{WEI01a}. We adopt their
rate coefficients for the recombination of $\Hp$, $\Hep$, $\cp$ and $\sip$. For
$\op$ and $\Dp$, we use the facts that $\op$ and $\Dp$ have almost the same 
ionization potential as $\Hp$ and that the ion arrival rate at the grain scales as
$m_{i}^{-1/2}$ (where $m_{i}$ is the mass of the ion in atomic mass units) to derive rates by
appropriately scaling the $\Hp$ rate. The \citeauthor{WEI01a} rates were all 
computed for Milky Way dust, and to adapt them for use in low metallicity 
gas, we again assume that they scale linearly with ${\rm Z}$, with the same 
caveats as before. 

\subsection{Photochemical rates}
\label{photochem}
In Table~\ref{tab:chem_gas_photo}, we list the cross-sections for all but two
of the photochemical reactions included in our model. The two exceptions 
are $\mHt$ photodissociation and $\hd$ photodissociation, which are 
caused by absorption in a large number of discrete spectral lines, and which are 
discussed separately in \S\ref{h2_photodiss} below.

Given the cross-section, $\sigma(E)$, the corresponding photochemical rate can be obtained 
from
\begin{equation}
R_{\rm photo} = 4\pi \int_{E_{0}}^{\infty} \frac{\sigma(E) I(E)}{E} e^{-\tau(E)} 
\left[1 + f(E)\right] {\rm d}E 
\end{equation}
where $E$ is the photon energy, $E_{0}$ is the energy threshold,
$I(E)$ and $\tau(E)$ are the mean intensity (in units of ${\rm eV} \: {\rm s^{-1}} \:
{\rm cm^{-2}} \: {\rm eV^{-1}} \: {\rm sr^{-1}}$) and optical depth 
for a photon of energy $E$ (both of which are problem dependent), and 
where $f(E)$ is a factor that accounts for the effects of secondary ionizations. It is 
generally a reasonable approximation to set $f(E) = 0$ unless the ionizing component 
of the radiation field is dominated by X-rays \citep{ABE97,GB03}. In the case that X-rays 
dominate, fits for $f(E)$ for $\mH$ and $\He$ photoionization as a function of the fractional 
ionization of the gas can be found in \citet{svs85} and \citet{dyl99}. The effects of secondary 
ionizations on the other processes listed here are generally negligible, owing
to the large abundances of neutral $\mH$ and $\He$ relative to all other species.

If the gas is optically thick at the $\He$ photoionization threshold, then an additional
process that must be taken into account is the photoionization of $\mH$ by the 
diffuse emission produced by $\Hep$ recombination. In the limit of high optical
depth, the on-the-spot approximation applies, and we can model this process as
a local ionization rate with a value \citep{os89}
\begin{equation}
R_{\rm pi} = \left[yk_{25, {\rm rr, A}} + (0.96-y) k_{25, {\rm rr, B}} + k_{25, {\rm di}} \right]
n_{\rm e} n_{\Hep} \: {\rm cm^{-3}} \: {\rm s^{-1}},
\end{equation}
where $y$ is given by
\begin{equation}
 y = \frac{n_{\mH} \sigma_{48}(E_{\rm th, He})}{n_{\mH} \sigma_{48}(E_{\rm th, He})
 + n_{\He} \sigma_{50}(E_{\rm th, He})},
 \end{equation}
where $E_{\rm th, He} = 24.6 \: {\rm eV}$ is the $\He$ ionization threshold. (Note that if 
$n_{\He}/n_{\mH} \simeq 0.08$, as is the case in primordial gas with a low fractional 
ionization and low $\mHt$ abundance, then $y \simeq 0.68$). If the on-the-spot 
approximation does not apply, then the radiative transfer of this diffuse emission must 
be modelled in some fashion. However, a discussion of appropriate techniques for 
doing so lies well beyond the scope of this paper. 

In some circumstances it may also be necessary to take account of the 
photodissociation of $\Hm$ and $\mHtp$ by photons produced by ionized 
hydrogen and helium (both recombination emission and bremsstrahlung). 
This process is generally important only in neutral gas close to a significant
volume of dense, ionized gas (e.g.\ in the neutral gas immediately 
surrounding an expanding ionization front). It is discussed in detail in 
\citet{GLO07} and so we do not discuss it further here.

Finally, we note that for each of the photoionization or photodissociation rates
listed in Table~\ref{tab:chem_gas_photo}, there is a corresponding photoheating
rate, given by
\begin{equation}
R_{\rm heat} =  4\pi \int_{E_{0}}^{\infty} \frac{\sigma(E) I(E)}{E} e^{-\tau(E)} 
(E-E_{0}) \eta(E-E_{0}) {\rm d}E.
\end{equation}
where $E-E_{0}$ is the energy of the primary photoelectron and 
$\eta(E-E_{0}) \leq 1$ gives the fraction of this energy converted to 
heat, which can be calculated using the results of \citet{svs85} or \citet{dyl99}.
In practice, photoheating from the photoionization of $\mH$ and $\He$ usually 
dominates over the other contributions by a wide margin.

\subsubsection{$\mHt$ and $\hd$ photodissociation}
\label{h2_photodiss}
Although the binding energy of $\mHt$ is only 4.48~eV, photons of this energy are not
able to dissociate $\mHt$ effectively, as the simplest dissociative transition --  excitation 
to the vibrational continuum of the ground state -- is strongly forbidden~\citep{fsd66}.
Transitions to the repulsive b$^{3}\Sigma_{u}^{+}$ state, the least energetic of the excited 
electronic states of $\mHt$, are also forbidden, and so photodissociation takes place 
primarily through excitation to the  Lyman (B$\,^{1}\Sigma^{+}_{u}$) or Werner 
(C$\,^{1}\Pi_{u}$) electronic states followed by radiative decay to the vibrational continuum 
of the ground state. As a number of vibrational levels are accessible in each excited state, 
photodissociation takes place through a number of discrete absorption lines, known as the 
Lyman and Werner band systems \citep{sw67}. 

In optically thin gas, the photodissociation rate can be written as
\begin{equation}
R_{\rm diss} = \sum_{v,J} R_{{\rm diss}, v,J} f_{v,J}
\end{equation}
where $f_{v,J}$ is the fraction of $\mHt$ molecules that have vibrational and rotational
quantum numbers $(v,J)$ in the electronic ground state, and $R_{{\rm diss}, v,J}$ is the 
photodissociation rate due to transitions out of $(v,J)$. The latter can be written as
\begin{equation}
 R_{{\rm diss},v,J} = \sum_{v^{\prime},J^{\prime}} \zeta_{v,J,v^{\prime},J^{\prime}}
 f_{{\rm diss},v^{\prime},J^{\prime}}, 
\end{equation}
where $\zeta_{v,J,v^{\prime},J^{\prime}}$ is the pumping rate from level $(v,J)$ in the 
electronic ground state to level $(v^{\prime},J^{\prime})$ in either the Lyman or Werner states, 
$f_{{\rm diss},v^{\prime},J^{\prime}}$ is the fraction of decays from $v^{\prime}, J^{\prime}$
which end in the vibrational continuum of the ground state (rather than back in some bound 
state), and where we sum over all accessible levels. Given appropriate molecular data, 
calculation of $R_{{\rm diss},v,J}$ is straightforward for each bound level $(v,J)$ in the
electronic ground state. To then calculate $R_{\rm diss}$, one also needs to know the level 
populations $f_{v,J}$. 

If we assume that the mean intensity $I(\nu) \equiv h I(E)$ is independent of energy, and that 
all of the $\mHt$ is in the $v=0, J=0$ level (i.e.\ the para-hydrogen ground state), then 
$R_{\rm diss}$ evaluates to 
\begin{equation}
 R_{\rm diss} = 1.38 \times 10^{9} I(\nu) \: {\rm s^{-1}}, \label{h2_thin_flat}
\end{equation}
where we have made use of molecular data taken from \citet{ab93a,ab93b} and \citet{ard00}. 
This expression remains a good approximation in the more general case that $I(\nu)$ is allowed 
to vary with frequency, provided that the variations are not too extreme and that we replace $I(\nu)$ in 
equation~\ref{h2_thin_flat} with $I(\bar{\nu})$, where $h\bar{\nu} = 12.87 \: {\rm eV}$ \citep{ABE97}:
\begin{equation}
 R_{\rm diss} = 1.38 \times 10^{9} I(\bar{\nu}) \: {\rm s^{-1}}. \label{h2_thin}
\end{equation}
Relaxing the assumption that all of the $\mHt$ has $J=0$ also makes little difference to 
$R_{\rm diss}$ \citep{gl01}. Vibrational excitation of the $\mHt$ makes a much larger difference 
\citep{sh78}, but at low gas densities we would expect the populations of the vibrational levels of 
$\mHt$ to be very small. Equation~\ref{h2_thin} therefore gives a reasonable estimate of the
optically thin $\mHt$ photodissociation rate within the regions of parameter space for which our
chemical model is valid. 

If enough $\mHt$ is present in the gas, then the Lyman-Werner lines can become optically
thick, leading to a reduction in the $\mHt$ photodissociation rate, an effect known as $\mHt$
self-shielding. If the gas is at rest, then the effects of $\mHt$ self-shielding can be treated 
quite accurately using the prescription of \citet{DRA96}. They parameterize the self-shielding
with a shielding function $f_{\rm sh}$, defined to be the ratio of the $\mHt$ photodissociation 
rate in self-shielded gas to the rate in optically thin gas. They demonstrate how to calculate 
$f_{\rm sh}$ as a function of the gas temperature and the $\mHt$ column density 
and also construct the following useful fitting function:
\begin{equation}
f_{\rm sh} = \frac{0.965}{(1 + x/b_{5})^{2}} + \frac{0.035}{(1 + x)^{1/2}}
\exp\left[-8.5 \times 10^{-4} (1+x)^{1/2}\right], \label{db_fsh}
\end{equation}
where $x = N_{\mHt} / 5 \times 10^{14} \: {\rm cm}^{-2}$, $N_{\mHt}$ is the $\mHt$
column density, $b_{5} = b / 10^{5} \: {\rm cm} \:{\rm s}^{-1}$ and $b$ is the Doppler 
broadening parameter. Although \citet{DRA96} assume a semi-inifinite slab
geometry in their models, their approach is easy to extend to more complicated
geometries. 

Unfortunately, their simple treatment breaks down in gas which is not at rest. Doppler
shifts due to the motions of the gas cause $\mHt$ in different regions to absorb at
slightly different wavelengths, and if these Doppler shifts are comparable to or larger
than the thermal linewidth of the gas (as will be the case in transonic or supersonic
gas respectively), then the effect is to reduce the amount of self-shielding that occurs.
An accurate treatment of $\mHt$ self-shielding in this regime probably requires one to
solve the full frequency-dependent transport equation, which cannot currently be done
in a computationally efficient manner within a three-dimensional hydrodynamics code.
Consequently, various different approximations have been used to study $\mHt$
photodissociation in this regime.

The simplest approach is to ignore self-shielding entirely \citep[see e.g.][]{MAC01,MAC03}.
This is a good approximation if the velocities in the gas are large and the $\mHt$
column densities are small, but otherwise will significantly overestimate the 
photodissociation rate. At the other extreme, one can ignore the effects of Doppler
shifts \citep[e.g.][]{YOS03,hi06}. This is a good approximation if the $\mHt$ column density
is sufficiently large ($N_{\mHt} > 10^{19} \: {\rm cm^{-2}}$) that the Lorentz wings of
the Lyman-Werner line profiles dominate the line widths, as in this case the line widths
will be much larger than any likely Doppler shifts within the molecular gas. On the other
hand, this approach will underestimate the true photodissociation rate when 
$N_{\mHt} < 10^{19} \: {\rm cm^{-2}}$, as is the case in many interesting low-metallicity 
systems.

Another approximation has recently been suggested by \citet{as07}. They use 
equation~(\ref{db_fsh}) to compute $f_{\rm sh}$, but adopt a value for $b$ that 
includes both a thermal contribution and one arising due to the velocity 
dispersion of the gas. In practice, this means that they treat their $\mHt$ as having
an effective $b$ equivalent to that in a purely thermal gas with 
$T = 10^{4} \: {\rm K}$. The accuracy of this approximation depends on the
correlation length of the velocity field. If this is small compared to the other length
scales of interest, then treating the velocity dispersion in this fashion is reasonable
and should give a fairly accurate result. On the other hand, if the velocity field is
dominated by large-scale bulk motions (such as infall into a protogalaxy), then 
this approximation will be significantly less accurate.

Finally, in \citet{GLO06} and in paper II, we use a local approximation in which only
the $\mHt$ within a single SPH smoothing length is assumed to contribute to the
shielding \citep[see also][for a grid-based version of this approach]{gm07a,gm07b}.
This fairly crude approximation is intended to take account of the fact that $\mHt$
close to a given point of interest is more likely to have only a small relative velocity
than gas a large distance away. It will generally underestimate the amount of 
self-shielding, but nevertheless represents an improvement over neglecting
self-shielding entirely. Aside from its inevitable inaccuracy, this approximation
also suffers from the disadvantage of being resolution dependent, as increasing the
number of SPH particles in the simulation will generally decrease all of the SPH
smoothing lengths and hence will cause a systematic increase in $f_{\rm sh}$.
On the other hand, it has the significant advantages of being computationally
efficient (as only local data is required) as well as being very easy to implement.

To sum up, a number of different approximate methods exist for treating $\mHt$
self-shielding in large numerical simulations, but none are entirely satisfying. 
Further work on this problem is definitely called for.

Turning now to $\hd$, we note that $\hd$ photodissociation in optically thin gas
can be treated in much the same way as $\mHt$ photodissociation. The 
necessary molecular data for $\hd$ can be found in \citet{ar06}, and the 
resulting photodissociation rate for a radiation field with a flat spectrum can be 
written as 
\begin{equation}
R_{\rm diss, \hd} = 1.5 \times 10^{9} I(\nu) \: {\rm s^{-1}},  \label{hd_flat}
\end{equation}
which is only $\sim 10\%$ larger than the $\mHt$ rate. Above an HD column
density $N_{\hd} \simeq 10^{13} \: {\rm cm^{-2}}$, self-shielding of the HD lines
significantly reduces the photodissociation rate. For a static gas, this process can 
again be modeled using the approach of \citet{DRA96}, although the same 
problems arise when one tries to extend this approach to a gas distribution which 
is not static. However, in the case of $\hd$,
we face an additional complication: if the $\hd$ column density is sufficiently 
high for HD self-shielding to be significant, then the $\mHt$ column density will
be very much larger (since even in significantly fractionated regions, one 
typically has an HD:$\mHt$ ratio of no more than about $10^{-3}$). Consequently,
the line widths of the $\mHt$ Lyman-Werner lines are not negligible, and some
degree of overlap between these lines and the HD absorption lines will occur.
Additionally, if the $\mHt$:H ratio is small, as will often be the case in the systems
of interest, then a significant HD column density implies a large neutral hydrogen 
column density, which means that absorption of radiation in the Lyman series lines
of atomic hydrogen must also be taken into account. 

These effects are difficult to
include accurately in a simple treatment of HD self-shielding and before attempting 
to do so it is reasonable to ask whether an accurate treatment of HD self-shielding
is really required. We argue that in many cases of interest it is not. Comparison 
of the rate at which $\hd$ is photodissociated in optically thin gas with the rate at
which it is destroyed by reaction 21 demonstrates that the latter dominates 
whenever $n_{\Hp} \simgreat 10^{-3} J_{21}(\bar{\nu})$, where $J_{21}(\bar{\nu})$ is 
the strength of the radiation field at $h\bar{\nu} = 12.87 \: {\rm eV}$ in units of
$10^{-21} \: {\rm ergs} \: {\rm s^{-1}} \: {\rm cm^{-2}} \: {\rm Hz^{-1}} \: {\rm sr^{-1}}$.
Photodissociation therefore dominates only when the UV field is strong or the 
proton number density is small. However, in either case, it is difficult to see how 
the large column densities of HD and $\mHt$ required for effective shielding could
be built up or maintained. Therefore, we suspect that for most applications, treating
the HD in the optically thin limit is probably sufficient, as in the conditions where this
approximation breaks down, photodissociation is unlikely to be important.

\section{Thermal processes}
\label{therm}
\subsection{Fine structure cooling}
\label{fscool}
As we do not include molecular coolants such as CO or ${\rm H_{2}O}$ in our chemical
model of metal-enriched gas, for the reasons outlined in \S\ref{chem_choice}, 
the main contribution that the metals make to the cooling of the gas is through fine structure 
line emission from neutral $\mC$, $\mO$ and $\mSi$ atoms and $\cp$ and $\sip$ ions. To
model this emission, we assume that the populations of all the electronically excited 
levels of these atoms and ions are negligible, an approximation which should be highly
accurate at the gas densities considered in this study. This assumption allows us to 
model $\cp$ and $\sip$ as two-level systems and $\mC$, $\mO$ and $\mSi$ as 
three-level systems, allowing us to compute their effects in a straightforward fashion. 
For a two-level ion, if we denote the ground state as level 0 and the excited state as
level 1, then the power radiated per unit volume can be written as
\begin{equation}
\Lambda = (A_{10} + B_{10} I_{10}) E_{10} n_{1},   \label{tl_cool}
\end{equation}
where $n_{1}$ is the number density of ions in level 1, $A_{10}$ is the Einstein
coefficient for spontaneous emission for the transition from level 1 to level 0,  
$B_{10}$ is the corresponding coefficient for stimulated emission,
$E_{10}$ is the energy of the transition, and $I_{10}$ is the mean specific intensity 
at the frequency of the transition. If  $I_{10} \neq 0$, then the ions will also absorb 
energy from the radiation field, at a rate
\begin{equation}
\Gamma = B_{01} I_{10} E_{10} n_{0}, 
\end{equation}
where $n_{0}$ is the number density of ions in level 0 and $B_{01}$ is the Einstein
coefficient for absorption from level 0 to level 1, which is related to $B_{10}$ by
$B_{01} = (g_{1}/g_{0}) B_{10}$, where $g_{0}$ and $g_{1}$ are the statistical 
weights of levels 0 and 1 respectively. The net loss of energy per unit time per
unit volume is therefore
\begin{eqnarray}
\Lambda^{\prime} & = & \Lambda - \Gamma \nonumber \\
& = & E_{10} \left( \{A_{10} n_{1} + B_{10} I_{10} [n_{1} - (g_{1} / g_{0}) n_{0}] \} \right),
\end{eqnarray}
and it will be seen that if 
\begin{equation}
n_{0} > \frac{A_{10} + B_{10} I_{10}}{(g_{1}/g_{0}) B_{10} I_{10}} n_{1},
\end{equation}
then the ions will absorb more energy than they emit, and so the gas will actually
gain energy.

To compute $\Lambda^{\prime}$, we need to know several pieces of atomic data
-- the values of $A_{10}$, $E_{10}$, $g_{0}$ and $g_{1}$, which are summarized 
for $\cp$ and $\sip$ in Table~\ref{fs_data} -- together with the values of 
$I_{01}$, $n_{0}$ and $n_{1}$. To compute $n_{0}$ and $n_{1}$, we assume that the levels 
are in statistical equilibrium, in which case:
\begin{equation}
(B_{01} I_{01} + C_{01}) n_{0} = (A_{10} + B_{10} I_{10} + C_{10}) n_{1},
\end{equation}
where $C_{01}$ and $C_{10}$ are the total rates of collisional excitation and 
de-excitation respectively. These are related by 
\begin{equation}
C_{01} = C_{10} \frac{g_{1}}{g_{0}} \expf{-}{E_{10}}{kT},
\end{equation}
and so once one is known, the other can be computed easily.  In Table~\ref{fs_coll_rates}, 
we list collisional de-excitation rates for collisions between $\cp$ or $\sip$ and various
possible collision partners such as $\mH$, $\mHt$ or $\me$. Given the number densities 
of these species, $C_{10}$ can be easily computed, since
\begin{equation}
C_{10} =  \sum_{k} q_{10, k} n_{k},
\end{equation}
where $q_{10,k}$ is the collisional de-excitation rate for a collision with chemical 
species $k$ with number density $n_{k}$. For $\cp$, we include the effects of collisions
with electrons, atomic hydrogen and molecular hydrogen (in both ortho and para forms).
The collision rate with $\Hp$ is negligible at the temperatures of interest due to the strong
Coloumb repulsion, and the abundances of the other chemical species included in our 
model are too small for them to be important collision partners. For $\sip$, we include 
only the effects of collisions with $\mH$ and $\me$, as rates for collisions with $\mHt$
are not available. However, provided that the $\mHt$ abundance is small compared to
the atomic hydrogen abundance, this is unlikely to be a major source of error.

Finally, to compute $I_{10}$, we assume that the only significant radiation field 
present at the infra-red and sub-millimeter wavelengths of the fine structure transitions
is the cosmic microwave background. In that case, $I_{10}$ is simply given by the 
value of the Planck function at the frequency of the transition for a radiation field with
temperature $T = T_{\rm CMB} = 2.726 (1+z) \: {\rm K}$.

For the three-level atoms ($\mC$, $\mO$ and $\mSi$), we use a very similar approach.
In this case, the power radiated per unit volume is
\begin{equation}
\Lambda = (A_{10} + B_{10} I_{10}) E_{10} n_{1} + (A_{20} + B_{20} I_{20}) E_{20} n_{2}
+ (A_{21} + B_{21} I_{21}) E_{21} n_{2},
\end{equation}
and the power absorbed per unit volume is
\begin{equation}
\Gamma = B_{01} I_{10} E_{10} n_{0} + B_{02} I_{20} E_{20} n_{0} + 
B_{12} I_{21} E_{21} n_{1}, 
\end{equation}
where 0, 1 and 2 denote the ground state and the two excited states respectively. The
level populations $n_{0}$, $n_{1}$ and $n_{2}$ are found by solving 
\begin{eqnarray}
(B_{01} I_{01} + C_{01} + B_{02} I_{02} + C_{02}) n_{0} & = &
(A_{10} + B_{10} I_{10} + C_{10}) n_{1} + (A_{20} + B_{20} I_{20} + C_{20}) n_{2}, \\
(B_{10} I_{10} + C_{10} + B_{12} I_{12} + C_{12}) n_{1} & = & 
(B_{01} I_{01} + C_{01}) n_{0} + (A_{21} + B_{21} I_{21} + C_{21}) n_{2},
\end{eqnarray}
with all symbols having their obvious meanings. To compute the total collisional 
excitation and de-excitation rates for carbon and oxygen we include the effects of 
collisions with ortho and para-$\mHt$, atomic hydrogen, protons and electrons.
For silicon, we include only the effects of collisions with $\mH$ and $\Hp$, as rates
for collisions with $\mHt$ or electrons do not appear to be available. The rates used 
are summarized in Table~\ref{fs_coll_rates}.

\subsection{Other coolants}
\label{cool:other}
Apart from fine structure emission, we also include in our thermal model several
other processes that can lead to the cooling of the gas. These are summarized in
Table~\ref{cool_other}, along with a reference to the source (or sources) from
which the associated cooling rate has been taken. In most cases, the rate itself
is also listed.

In hot, ionized gas, cooling is dominated by electron impact excitation of atomic
hydrogen (Lyman-$\alpha$ cooling), atomic helium and $\Hep$. Excitation of 
atomic helium occurs from both the $1^1$S ground state and the $2^3$S 
metastable state. In common with previous authors, we assume that the 
population of the $2^3$S state is set by the balance between radiative 
recombination to triplet states and radiative decay to the ground state.
One consequence of this assumption is that the number density of $\He$ 
atoms in the  $2^3$S state, $n_{{\rm He} (2^3{\rm S})}$, is proportional to
the product of the number densities of free electrons and of $\Hep$, i.e.\
 $n_{{\rm He} (2^3{\rm S})} \propto n_{\rm e} n_{\Hep}$, which means that
 the cooling rate from metastable helium scales as 
 $n_{{\rm He} (2^3{\rm S})} n_{\rm e} \propto n_{\rm e}^{2} n_{\Hep}$. 
To model cooling from $\mH$, the $\He$ metastable state, and $\Hep$, we use 
rates from \citet{CEN92}, which themselves were based on earlier work by \citet{BLA81}.
To model cooling from the $\He$ ground state, we use our own fit to the data of
\citet{BBFT00}. 

A number of other processes are of importance in ionized gas. We include cooling
due to collisional ionization of atomic hydrogen and atomic helium, the gas phase
recombination of $\Hp$ and $\Hep$, ionic recombination on dust grains, Compton 
scattering of CMB photons by free electrons (Compton cooling), and thermal 
bremsstrahlung. The rates adopted for all of these processes are summarized in
Table~\ref{cool_other}. As the chemical and thermal model presented here is not
designed to be used for the study of very hot gas, we do not include processes 
involving  ${\rm He^{++}}$ (although rates for these processes can be found in 
\citealt{CEN92}).

In neutral gas, all of the aforementioned processes become ineffective. In cool,
neutral gas, most of the cooling comes from $\mHt$ or from the fine structure 
lines of carbon, oxygen and silicon, which have already been discussed above. 
Cooling from $\mHt$ is treated in our model through use of the cooling function of 
\citet{BOU99}, which we have extended to temperatures below $100 \: {\rm K}$ by 
assuming that only the $J = 2 \rightarrow 0$ and  $J = 3 \rightarrow 1$ transitions 
contribute significantly to the cooling rate. \citet{BOU99} tabulate the $\mHt$ cooling 
rate as a function of temperature, density,
$\mH$:$\mHt$ ratio and ortho:para ratio. For simplicity, in our implementation we
do not track the evolution of the $\mHt$ ortho:para ratio, instead keeping it fixed at
 3:1, but we note that variations in this ratio are unlikely to significantly affect the 
$\mHt$ cooling rate at  temperatures at which it contributes significantly to the total 
cooling rate (see, for instance, figure~5 in \citealt{BOU99}). A comparison of the 
\citeauthor{BOU99} $\mHt$ cooling rate with various other rates that have been 
used in the literature is given in paper II (Figure~1).

At low temperatures ($T \simless 200 \: {\rm K}$),  $\hd$ cooling becomes 
increasingly important and can dominate the total cooling rate if sufficient 
fractionation occurs \citep{GP98}. To model $\hd$ cooling, we use the recent 
cooling function of \citet{lna05}. They provide a complicated fit as a value of 
temperature and density that is valid over a wide range of both. For further details, 
the interested reader should consult their paper.

We include two final processes that can become important in some 
circumstances. Cooling due to $\mHt$ collisional dissociation is
modelled under the assumption that each dissociation removes 
$\sim$4.48~eV of thermal energy from the gas. It is an effective source
of cooling only at temperatures above a few thousand Kelvin. In practice,
the $\mHt$ abundance in zero metallicity or low metallicity gas is frequently
too small for this process to be important. 

We also include the effects of energy transfer from the gas to the dust grains 
(if present), using a rate from \citet{HOL89}. At solar metallicity, this becomes
important at a density $n \sim 10^{4} \: {\rm cm^{-3}}$, but at lower metallicities,
it does not dominate until significantly larger densities are reached 
\citep[see e.g.][]{OMU05}.

\subsection{Heating}
\label{heat:other}
We include in our model the effects of several processes that can heat the gas.
Most of these processes operate only if a radiation background is present. The
first of these is photoelectric emission from dust grains. This operates as follows: 
photons that interact with dust grains
can cause the ejection of energetic electrons from the grain if the photon energy
exceeds the work function of the grain. As the energy carried by the electrons is
quickly thermalized, this leads to the heating of the gas. This process is of great
importance in the local ISM and has been examined in detail by a number of
authors \citep[e.g.][]{BAK94,WOL95,WEI01b,RAE04}.

To accurately compute the effects of photoelectric emission, we need to know
the grain size distribution and the composition of the grains. However, as 
discussed previously, large uncertainties exist concerning the properties of
grains in low metallicity protogalactic gas. We therefore make the same 
assumption here as we did in our treatment of grain surface chemistry, i.e.\
that the dust has the same properties as Milky Way dust, but has an abundance
that is reduced by a factor $({\rm Z}/{\rm Z_{\odot}})$. This assumption allows
us to use the following expression for the photoelectric heating rate, taken from 
\citet{WOL95}:
\begin{equation}
\Gamma_{\rm pe} = 1.3 \times 10^{-24} \epsilon \, G 
\left(\frac{{\rm Z}}{{\rm Z_{\odot}}}\right) n \: {\rm erg} \: {\rm s^{-1}} \: {\rm cm^{-3}},
\end{equation}
where $G \simeq 0.01 J_{21}$ is a measure of the radiation energy density between 
$6 \: {\rm eV}$  and $13.6 \: {\rm eV}$ relative to the \citet{habing68} field, and 
where $\epsilon$ is the photoelectric heating efficiency, given by
\begin{equation}
\epsilon = \frac{4.9 \times 10^{-2}}{1.0 + 4.0 \times 10^{-3} \tilde{\psi}^{0.73}} +
\frac{3.7 \times 10^{-2} T_{4}^{0.7}}{1.0 + 2.0 \times 10^{-4} \tilde{\psi}},
\end{equation}
where $T_{4} = T / 10^{4} \: {\rm K}$. The parameter $\tilde{\psi}$ that controls
the photoelectric heating efficiency is given by $\tilde{\psi} = G \sqrt{T} / 0.5 n_{\rm e}$
\citep{WOL03}; note that this differs by a factor of 2 from the parameter $\psi$ 
introduced in \S\ref{surface_chem}. For small $\tilde{\psi}$, most grains 
are negatively charged and $\epsilon \simeq 4.9 \times 10^{-2} + 
3.7 \times 10^{-2} T_{4}^{0.7}$. On the other hand, for large 
$\tilde{\psi}$, most grains are positively charged, and $\epsilon$ is small, as it is 
difficult for photons with energies $E < 13.6 \: {\rm eV}$ to dislodge further 
electrons from the grains. 

A second source of radiative heating is the photodissociation of $\mHt$. Our 
treatment of photodissociation heating follows \citet{BLA77}: we assume that 
each photodissociation deposits $0.4 \: {\rm eV}$ of themal energy into the gas.
As well as photodissociating some of the $\mHt$, an ultraviolet background
will also produce vibrationally excited $\mHt$ via radiative pumping of the
excited levels. In dense gas, this pumping leads to heating as most of
the excited molecules undergo collisional de-excitation. We include the 
effects of radiative pumping by adopting a pumping rate that is 8.5 times larger 
than the photodissociation rate \citep{DRA96}, and assuming that each excitation 
transfers an average of $2 \, (1 + n_{\rm cr}/n)^{-1} \: {\rm eV}$ to the gas 
\citep{BUR90}, where $n_{\rm cr}$ is the critical density at which collisional 
de-excitation of vibrationally excited $\mHt$ occurs at the same rate as radiative 
de-excitation. Our value for $n_{\rm cr}$ is a weighted harmonic mean of the 
value for $\mHt$-$\mH$ collisions given by \citet{LEP83}, reduced by a factor 
of ten as advised by \citet{MAR96}, and the value for $\mHt$-$\mHt$ collisions 
given by \citet{SHA87}. 

Heating due to the photoionization of $\mH$ or $\He$ has already been discussed
in \S\ref{photochem} and we do not discuss it further here. As is also discussed in that 
section, we do not include heating due to the photodetachment of $\Hm$, photodissociation 
of $\mHtp$ or photoionization of $\mC$, $\mO$, $\mSi$ or $\sip$, as the contribution 
from these processes is not significant.

We also include the effects of heating due to $\mHt$ formation. The formation of 
an $\mHt$ molecule via reaction 2 releases 3.53~eV of energy, while formation 
via reaction 4 releases 1.83~eV, and formation on a grain surface (reaction 60) 
releases 4.48~eV.  We assume that essentially all of this energy goes into rotational 
and vibrational excitation of the resulting $\mHt$ molecule, and hence is radiated
away at low gas densities and is converted by collisional de-excitation into heat 
at high gas densities. 

Finally, we include heating due to the ionization of the gas by cosmic rays. 
Following \citet{gl78}, we assume that every ionization deposits 20~eV of heat
in the gas, and so derive a heating rate that scales with the total cosmic ray
ionization rate of the gas (i.e.\ the sum of the ionization rates for the various 
individual chemical species, weighted by the fractional abundances of those
species). Since considerable uncertainty exists concerning the value of the cosmic 
ray ionization rate in the local interstellar medium \citep[see the discussion in][]{mcc03},
let alone concerning the appropriate rate to use in high-redshift protogalaxies, 
a more detailed treatment does not appear to be warranted at this time.

\section{Applications}
\label{appl}
The chemical network and thermal model described in the preceding sections
have a number of potential applications. One of the more obvious applications
is the study of the cooling and gravitational collapse of gas in low metallicity
protogalaxies. As previously noted, the density range for which our model is
valid corresponds to a wide range of cosmological overdensities. It can 
therefore be used to study the thermal and chemical evolution of the majority
of the gas within a given protogalaxy. For example, suppose we model the
gas distribution within a $z=20$ halo as a singular isothermal sphere with
mean number density within the virial radius $\bar{n} = 0.4 \: {\rm cm^{-3}}$
(a reasonable zeroth-order approximation for a protogalactic halo with a 
mean overdensity $\delta = 200$; see e.g.\ \citealt{ABE02}). In this case,
only $\sim$4\% of the gas within the halo has a density $n > 100 \: {\rm cm^{-3}}$
at which our model may break down. Therefore, although our model is of limited
usefulness for studying gravitational fragmentation and star formation within
this dense central region, it does allow one to study many other important
problems, such as how the minimum protogalactic mass at which cooling 
becomes effective is affected by the presence of metals, or how the gas 
responds to the presence of an ultraviolet background. We use a version
of this model to address some of these questions in paper II for the case of 
small protogalactic halos within initially ionized regions. 

A second possible application is the study of the evolution of the ISM in high
redshift, metal-poor dwarf galaxies. In this case, much of the gas involved is
often gravitational stable, and so the characteristic timescales are longer
than the gravitational free-fall time. This limits the applicability of the model
to lower densities than in the case of gravitationally collapsing protogalaxies.
However, even if we adopt a relatively long characteristic timescale
$t_{\rm char} = 100 \: {\rm Myr}$, corresponding to the sound crossing time of
a $1 \: {\rm kpc}$ disk at a sound speed of $10 \: {\rm km} \: {\rm s^{-1}}$, the
model remains valid for densities up to $n \sim 5 \: {\rm cm^{-3}}$ even in the
least auspicious case (${\rm Z} = 0.1 \: {\rm Z_{\odot}}$ gas, with no significant
UV radiation field present). Our model is therefore well-suited for use in the 
study of the evolution of the warm neutral component of the ISM in such 
galaxies and will in some cases also be useful in the study of the cold neutral
component. However, the reader is reminded that our model does not
treat hot ($T \gg 10^{4} \: {\rm K}$), highly-ionized gas, and cannot be used 
to study the effects of stellar feedback (\hii regions, supernovae, etc.) unless
coupled with some existing model capable of treating this hot gas 
(see e.g. \citealt{sd93}).

As well as studying the chemical evolution of gas {\em within} dwarf
galaxies and protogalaxies,
we can also use it to study the chemical evolution of gas {\em between}
galaxies, i.e.\ the intergalactic medium (IGM). One important reason to 
do so is the fact that several different obervational techniques have been
suggested that may allow one to probe the thermal and chemical state of 
this gas. \citet{oh02} has suggested that if large regions of the IGM remain
neutral after the turn-on of the first observable ionizing sources, then 
atomic oxygen in the IGM may be detectable through its UV absorption, 
as it will produce an \oi forest analogous to the $z < 6$ \hi
Lyman-$\alpha$ forest. \citet{fl03} also argue that a substantial
fraction of the metals in high-redshift Galactic winds are likely to be in
low-ionization states and may be observable in absorption. 
Metals in the high redshift IGM may also leave detectable imprints in the
fluctuation spectrum of the cosmic microwave background \citep{bhs04,hrs06}.
Finally, \citet{hmh07} have recently suggested that fine structure emission 
from atomic oxygen may also produce a detectable spectral distortion of the 
CMB if the excited fine-structure levels can be populated by radiative pumping 
via the \oi Balmer-$\alpha$ transition.
Accurate modelling of many of these effects requires accurate modelling of
the temperature and chemical make-up of the intergalactic gas, and hence
a model such as that presented here.

\acknowledgments
We thank R. Klessen and M.-M. Mac Low for useful discussions, and the
anonymous referee for valuable feedback on earlier drafts of this paper.
We also thank R.-S. Ciobanu for pointing out a serious typographical
error in an earlier version of this paper.
S.C.O.G. acknowledges support from NSF grant AST03-07793 during the early
phases of this work. A.K.J. acknowledges support from the Emmy Noether 
Program of the Deutsche Forschungsgemeinschaft  (grant no.\ KL1358/1).

\begin{deluxetable}{llllc}
\tablewidth{0pt}
\tabletypesize{\footnotesize}
\tablecaption{List of the collisional gas-phase reactions in our
chemical model. \label{tab:chem_gas_coll}}
\tablehead{No.\  & Reaction & Rate coefficient $({\rm cm}^{3} \: {\rm s}^{-1})$ &
& Ref.\ } 
\startdata
& & & & \\
1 & $\mH  + \me  \rightarrow \Hm + \gamma$ & 
$k_{1} = {\rm dex}[-17.845 + 0.762 \log{T} + 0.1523 (\log{T})^{2}$ & & 1 \\
& & $\phantom{k_{1}= {\rm dex}[} \mbox{} - 0.03274 (\log{T})^{3}] $ &
$T \le 6000 \: {\rm K}$ & \\
& & & & \\
& & $ \phantom{k_{1}} = {\rm dex}[-16.420 + 0.1998 (\log{T})^{2}$ & & \\ 
& & $ \phantom{k_{1} = {\rm dex}} \mbox{}-5.447  \times 10^{-3}  (\log{T})^{4}$ & & \\ 
& & $ \phantom{k_{1} = {\rm dex}} \mbox{}+ 4.0415 \times 10^{-5} (\log{T})^{6}]$ 
& $T > 6000 \: {\rm K}$ & \\
& & & & \\
\hline
& & & & \\
2 & $\Hm  + \mH  \rightarrow \mHt + \me$ &
$k_{2} = 1.5 \times 10^{-9}$ & $T \le 300 \: {\rm K}$& 2 \\
& & & & \\
& & $\phantom{k_{2}} = 4.0 \times 10^{-9} T^{-0.17}$ & $T > 300 \: {\rm K}$ & \\
& & & & \\
\hline
& & & & \\
3 & $\mH  + \Hp  \rightarrow \mHtp + \gamma$ &
$k_{3} = {\rm dex}[-19.38 - 1.523 \log{T} $ & & 3 \\
& & $\phantom{k_{3}} \mbox{}+1.118 (\log{T})^{2}  - 0.1269 (\log{T})^{3}]$ & & \\
 & & & & \\
 \hline
& & & & \\
4 & $\mH + \mHtp \rightarrow \mHt + \Hp$ & $k_{4} = 6.4 \times 10^{-10}$ & & 4 \\
 & & & & \\
 \hline
& & & & \\
5 & $\Hm  + \Hp  \rightarrow \mH + \mH$ & 
$k_{5} = 5.7 \times 10^{-6} T^{-0.5}   +  6.3 \times 10^{-8} $ & & 5 \\
& & $\phantom{k_{5}} \mbox{} - 9.2 \times 10^{-11} T^{0.5}  +  4.4 \times 10^{-13} T$ & \\
& & & & \\
\hline
& & & & \\
 6 & $\mHtp + \me \rightarrow \mH + \mH$ & $k_{6} = 
 1.0\times 10^{-8}$ &  $T \le 617 \: {\rm K}$ & 6 \\
 & & & & \\
 & & $\phantom{k_{6}} = 1.32 \times 10^{-6} T^{-0.76}$ & $T > 617 \: {\rm K}$ & \\
 & & & & \\
 \hline
& & & & \\
7 & $\mHt + \Hp  \rightarrow \mHtp + \mH$ &
$k_{7} = [- 3.3232183 \times 10^{-7}$ & & 7 \\
 & & $\phantom{k_{7}=}  \mbox{} + 3.3735382 \times 10^{-7}  \ln{T}$ & & \\
 & & $\phantom{k_{7}=}  \mbox{} - 1.4491368 \times 10^{-7}  (\ln{T})^2$ & & \\
 & & $\phantom{k_{7}=}  \mbox{} + 3.4172805 \times 10^{-8}  (\ln{T})^3$ & & \\
 & & $\phantom{k_{7}=}  \mbox{} - 4.7813720 \times 10^{-9}  (\ln{T})^4$ & & \\
 & & $\phantom{k_{7}=}  \mbox{} + 3.9731542 \times 10^{-10} (\ln{T})^5$ & & \\
 & & $\phantom{k_{7}=}  \mbox{}  - 1.8171411 \times 10^{-11}  (\ln{T})^6$ & & \\
 & & $\phantom{k_{7}=}  \mbox{}  + 3.5311932 \times 10^{-13} (\ln{T})^7 ]$ & & \\
 & & $\phantom{k_{7}=} \mbox{} \times \exp \left(\frac{-21237.15}{T} \right)$ & & \\
 & & & & \\
\hline
& & & & \\
8 & $\mHt + \me  \rightarrow  \mH + \mH +  \me$ &
$ k_{8} = 3.73 \times 10^{-9} T^{0.1121} \exp\left(\frac{-99430}{T}\right) $
& & 8 \\
& & & & \\
\hline
& & & & \\
9 & $\mHt + \mH  \rightarrow  \mH + \mH + \mH$  & 
$ k_{9} = 6.67 \times 10^{-12} T^{1/2} \exp \left[-(1+ \frac{63590}{T}) \right]$ & & 9 \\
 & & & & \\
 \hline
& & & & \\
10 & $\mHt + \mHt \rightarrow  \mHt + \mH + \mH$ & 
$k_{10} = \frac{5.996 \times 10^{-30} T^{4.1881}}{(1.0 + 6.761 \times 10^{-6} T)^{5.6881}}
 \exp \left(-\frac{54657.4}{T} \right)$ & & 10 \\
 & & & & \\
  \hline
& & & & \\
11 & $\mH  + \me  \rightarrow \Hp + \me + \me$ & 
$k_{11} =  \exp[-3.271396786 \times 10^{1}$ & & 11 \\
& & $\phantom{k_{11}=}  \mbox{}  + 1.35365560 \times 10^{1} \ln T_{\rm e}$ & & \\
& & $\phantom{k_{11}=}  \mbox{}  - 5.73932875 \times 10^{0} (\ln T_{\rm e})^{2}$ & & \\
& & $\phantom{k_{11}=}  \mbox{}  + 1.56315498 \times 10^{0} (\ln T_{\rm e})^{3}$ & & \\
& & $\phantom{k_{11}=}  \mbox{}  -  2.87705600 \times 10^{-1} (\ln T_{\rm e})^{4}$ & & \\
& & $\phantom{k_{11}=}  \mbox{}  + 3.48255977 \times 10^{-2} (\ln T_{\rm e})^{5}$ & & \\
& & $\phantom{k_{11}=}  \mbox{}   - 2.63197617 \times 10^{-3} (\ln T_{\rm e})^{6}$ & & \\ 
& & $\phantom{k_{11}=}  \mbox{}  + 1.11954395\times 10^{-4} (\ln T_{\rm e})^{7}$ & & \\
& & $\phantom{k_{11}=}  \mbox{}   -  2.03914985 \times 10^{-6} (\ln T_{\rm e})^{8}]$ & & \\
& & & & \\
 \hline
 & & & & \\
12 & $\mD + \me \rightarrow \Dp + \me + \me$ & $k_{12} = k_{11}$ & & --- \\
& & & & \\
 \hline
& & & & \\
13 & $\Hp  + \me  \rightarrow \mH +  \gamma$ & 
 $k_{13, {\rm A}} = 1.269 \times 10^{-13} \left(\frac{315614}{T}\right)^{1.503}$ & Case A & 12 \\
 & & $\phantom{k_{13}=} \mbox{} \times
 [1.0+ \left(\frac{604625}{T}\right)^{0.470}]^{-1.923} $ & & \\
& & $k_{13, {\rm B}} = 2.753 \times 10^{-14} \left(\frac{315614}{T}\right)^{1.500}$ & Case B & 12 \\
 & & $\phantom{k_{13}=} \mbox{} \times
 [1.0+ \left(\frac{115188}{T}\right)^{0.407}]^{-2.242} $ & & \\
 & & & & \\
\hline
 & & & & \\
14 & $\Dp + \me \rightarrow \mD + \gamma$ & $k_{14} = k_{13}$ & & --- \\
 & & & & \\
\hline
 & & & & \\
15 & $\Hm  + \me  \rightarrow \mH + \me + \me$ &
$ k_{15}  = \exp [-1.801849334 \times 10^{1}$ & & 11 \\
& & $\phantom{k_{15}=}  \mbox{} + 2.36085220 \times 10^{0} \ln T_{\rm e}$ & & \\
& & $\phantom{k_{15}=} \mbox{} - 2.82744300 \times 10^{-1} (\ln T_{\rm e})^{2}$ & & \\
& & $\phantom{k_{15}=} \mbox{}  +1.62331664\times 10^{-2} (\ln T_{\rm e})^{3}$ & & \\
& & $\phantom{k_{15}=} \mbox{} -3.36501203 \times 10^{-2} (\ln T_{\rm e})^{4}$ & & \\
& & $\phantom{k_{15}=} \mbox{}   +1.17832978\times 10^{-2} (\ln T_{\rm e})^{5}$ & & \\ 
& & $\phantom{k_{15}=} \mbox{}  -1.65619470\times 10^{-3} (\ln T_{\rm e})^{6}$ & & \\ 
& & $\phantom{k_{15}=} \mbox{}   +1.06827520\times 10^{-4} (\ln T_{\rm e})^{7}$ & & \\
& & $\phantom{k_{15}=} \mbox{}  -2.63128581\times 10^{-6} (\ln T_{\rm e})^{8} ]$ & & \\
 & & & & \\
  \hline
& & & & \\
16 & $\Hm  + \mH  \rightarrow  \mH + \mH +  \me$ &
$k_{16} = 2.5634 \times 10^{-9} T_{\rm e}^{1.78186}$ & $ T_{\rm e} \le 0.1 \: \rm{eV}$ & 11 \\
& & & & \\
& & $\phantom{k_{16}} = \exp[-2.0372609 \times 10^{1}$ & & \\
& & $\phantom{k_{16}=}  \mbox{}+1.13944933 \times 10^{0} \ln T_{\rm e}$ & & \\
& & $\phantom{k_{16}=} \mbox{}-1.4210135 \times 10^{-1} (\ln T_{\rm e})^{2}$ & & \\
& & $\phantom{k_{16}=} \mbox{}+8.4644554 \times 10^{-3} (\ln T_{\rm e})^{3}$ & & \\
& & $\phantom{k_{16}=}  \mbox{}-1.4327641 \times 10^{-3} (\ln T_{\rm e})^{4}$  & & \\
& & $\phantom{k_{16}=}  \mbox{}+2.0122503 \times 10^{-4} (\ln T_{\rm e})^{5}$ & & \\
& & $\phantom{k_{16}=}  \mbox{}+8.6639632 \times 10^{-5} (\ln T_{\rm e})^{6}$ & & \\
& & $\phantom{k_{16}=}  \mbox{}-2.5850097 \times 10^{-5} (\ln T_{\rm e})^{7}$ & & \\
& & $\phantom{k_{16}=} \mbox{}+ 2.4555012\times 10^{-6} (\ln T_{\rm e})^{8}$ & & \\
& & $\phantom{k_{16}=} \mbox{} -8.0683825\times 10^{-8} (\ln T_{\rm e})^{9}]$ & 
$T_{\rm e} > 0.1 \: \rm{eV}$ & \\ 
& & & & \\
\hline
& & & & \\
17 & $\Hm + \Hp   \rightarrow \mHtp + \me$ & 
$k_{17}= 6.9\times 10^{-9}  T^{-0.35}$  & $T \le 8000 \: {\rm K}$ & 13 \\
& & & & \\
 & & $\phantom{k_{17}} = 9.6 \times 10^{-7} T^{-0.90}$ & $T > 8000 \: {\rm K}$ & \\
& & & & \\
\hline
& & & & \\
18 & $\mH + \Dp \rightarrow \mD + \Hp$ & $k_{18} = 2.06 \times 10^{-10} T^{0.396}
\expf{-}{33}{T}$ & & 14 \\
& & $\phantom{k_{18}=} \mbox{} + 2.03 \times 10^{-9} T^{-0.332}$ & & \\
& & & & \\
\hline
& & & & \\
19 & $\mD + \Hp \rightarrow \mH + \Dp$ & 
$k_{19} = 2.0 \times 10^{-10} T^{0.402} \expf{-}{37.1}{T}$ & $T \le 2 \times 10^{5} \: {\rm K}$ & 14 \\
& & $\phantom{k_{19} = } \mbox{} - 3.31 \times 10^{-17} T^{1.48}$ &  & \\
& & & & \\
& & $\phantom{k_{19}} = 3.44 \times 10^{-10} T^{0.35}$ & $T > 2 \times 10^{5} \: {\rm K}$ & \\
& & & & \\
\hline
& & & & \\
20 & $\mHt + \Dp \rightarrow \hd + \Hp$ &
$k_{20} = \left[0.417 + 0.846 \log{T} - 0.137 (\log{T})^{2} \right] \times 10^{-9}$ & & 15 \\
& & & & \\
\hline
& & & & \\
21 & $\hd + \Hp \rightarrow \mHt + \Dp$ & $k_{21} = 1.1 \times 10^{-9} \expf{-}{488}{T}$ & & 15 \\
& & & & \\
\hline
& & & & \\
22 & $\mHt + \mD \rightarrow \hd + \mH$ & 
$k_{22} = 1.69 \times 10^{-10} \expf{-}{4680}{T}$ & $T \leq 200 \: {\rm K}$ & 16 \\
& & & & \\
& & $\phantom{k_{22}} =1.69 \times 10^{-10} \exp\left(-\frac{4680}{T} + 
\frac{198800}{T^{2}}\right)$ & $T > 200 \: {\rm K}$ & \\
& & & & \\
\hline
& & & & \\
23 & $\hd + \mH \rightarrow \mD + \mHt$ & 
$k_{23} = 5.25 \times 10^{-11} \expf{-}{4430}{T}$ & $T \leq 200 \: {\rm K}$ & 17 \\
& & & & \\
& & $\phantom{k_{23}} = 5.25 \times 10^{-11} \exp\left(-\frac{4430}{T} + \frac{173900}{T^{2}}\right)$
& $T > 200 \: {\rm K}$ & \\
& & & & \\
\hline
& & & & \\
24 & $\He + \me \rightarrow \Hep + \me + \me$ &
$k_{24} =  \exp[-4.409864886 \times 10^{1}$ & & 11 \\
& & $\phantom{k_{24} = } \mbox{}  + 2.391596563 \times 10^{1} \ln T_{\rm e}$ & & \\
& & $\phantom{k_{24} = } \mbox{}  - 1.07532302 \times 10^{1} (\ln T_{\rm e})^{2}$ & & \\
& & $\phantom{k_{24} = } \mbox{}  + 3.05803875 \times 10^{0} (\ln T_{\rm e})^{3}$ & & \\
& & $\phantom{k_{24} = } \mbox{}  -  5.6851189 \times 10^{-1} (\ln T_{\rm e})^{4}$ & & \\
& & $\phantom{k_{24} = } \mbox{}  + 6.79539123 \times 10^{-2} (\ln T_{\rm e})^{5}$ & & \\
& & $\phantom{k_{24} = } \mbox{}   - 5.0090561 \times 10^{-3} (\ln T_{\rm e})^{6}$ & & \\ 
& & $\phantom{k_{24} = } \mbox{}  + 2.06723616\times 10^{-4} (\ln T_{\rm e})^{7}$ & & \\
& & $\phantom{k_{24} = } \mbox{}   -  3.64916141 \times 10^{-6} (\ln T_{\rm e})^{8}]$ & & \\
& & & & \\
\hline
& & & & \\
25 & $\Hep + \me \rightarrow \He + \gamma$ & 
$k_{25, {\rm rr, A}} = 10^{-11} T^{-0.5} \left[12.72 - 1.615 \log{T} \right. $ & Case A & 18  \\
& & $\left. \phantom{k_{25, {\rm rr, A}} = } \mbox{} - 0.3162 (\log{T})^{2} + 0.0493 (\log{T})^{3}\right]$ & & \\
& & & & \\
& & $k_{25, {\rm rr, B}} = 10^{-11} T^{-0.5} \left[11.19 - 1.676 \log{T} \right. $ & Case B & 18  \\
& & $\left. \phantom{k_{25, {\rm rr, A}} = } \mbox{} - 0.2852 (\log{T})^{2} + 0.04433 (\log{T})^{3} \right]$ & & \\
& & & & \\
& & $k_{25, {\rm di}} = 1.9 \times 10^{-3} T^{-1.5} \expf{-}{473421}{T}$ & & \\
& & $\phantom{k_{25, {\rm di}} = } \mbox{} \times \left[1.0 + 0.3 \expf{-}{94684}{T} \right] $ & & 19 \\
& & & & \\
\hline
& & & & \\
26 & $\Hep + \mH \rightarrow \He + \Hp$ & 
$k_{26} = 1.25 \times 10^{-15} \left(\frac{T}{300}\right)^{0.25}$ & & 20 \\
& & & & \\
\hline
& & & & \\
27 & $\He + \Hp \rightarrow \Hep + \mH$ &
$k_{27} = 1.26 \times 10^{-9} T^{-0.75} \expf{-}{127500}{T}$ & $ T \leq 10000 \: {\rm K}$ & 21 \\
& & & & \\
& & $\phantom{k_{27}} = 4.0 \times 10^{-37} T^{4.74}$ & $T > 10000 \: {\rm K}$ &  \\
& & & & \\
\hline
& & & & \\
28 & $\Hep + \mD \rightarrow \He + \Dp$ & $k_{28} = k_{26}$ & & --- \\
& & & & \\
\hline
& & & & \\
29 & $\He + \Dp \rightarrow \Hep + \mD$ &  $k_{29} = k_{27}$ & & --- \\
& & & & \\
\hline
& & & & \\
30 & $\cp + \me \rightarrow \mC  + \gamma$ &
$k_{30} = 4.67 \times 10^{-12}  \left(\frac{T}{300}\right)^{-0.6}$ & $T \le 7950 \: {\rm K}$ & 22 \\
& & & & \\
& & $\phantom{k_{30} } =1.23 \times 10^{-17}  \left(\frac{T}{300}\right)^{2.49} 
\exp \left(\frac{21845.6}{T} \right)$ & $ 7950 \: {\rm K} < T \le 21140 \: {\rm K}$ & \\
& & & & \\
& & $\phantom{k_{30}} = 9.62 \times 10^{-8} \left(\frac{T}{300}\right)^{-1.37} 
 \exp \left(\frac{-115786.2}{T} \right)$ & $T > 21140 \: {\rm K}$ & \\
 & & & & \\
 \hline
& & & & \\
31 & $\sip + \me \rightarrow \mSi + \gamma$ &
$k_{31} =  7.5 \times 10^{-12} \left(\frac{T}{300}\right)^{-0.55}$  & $T \le 2000 \: {\rm K}$ & 23 \\
& & & & \\
& & $\phantom{k_{31}}= 4.86 \times 10^{-12} \left(\frac{T}{300}\right)^{-0.32}$ & 
$2000 \: {\rm K} < T \le 10^{4} \: {\rm K}$ & \\
& & & & \\
& & $\phantom{k_{31}}= 9.08 \times 10^{-14} \left(\frac{T}{300}\right)^{0.818}$ & 
$T > 10^{4} \: {\rm K}$ & \\
 & & & & \\
  \hline
& & & & \\ 
 32 & $\op + \me  \rightarrow \mO + \gamma$ &
$k_{32} = 1.30 \times 10^{-10} T^{-0.64}$ &  $T \le 400 \: {\rm K}$ & 24 \\
& & & & \\ 
& & $\phantom{k_{32}} = 1.41 \times 10^{-10} T^{-0.66} + 7.4 \times 10^{-4}  T^{-1.5}$ & & \\
& & $\phantom{k_{32}=} \mbox{} \times \exp \left(-\frac{175000}{T}\right) [1.0 + 0.062 \times 
\exp \left(-\frac{145000}{T}\right) ]$ & $T > 400 \: {\rm K}$ & \\
 & & & & \\
   \hline
& & & & \\ 
33 & $\mC  + \me  \rightarrow \cp  + \me + \me$ & 
$k_{33} = 6.85 \times 10^{-8} (0.193 + u)^{-1} u^{0.25} e^{-u}$ & $u = 11.26 / T_{\rm e}$ & 25 \\
 & & & & \\
\hline
 & & & & \\
34 & $\mSi + \me  \rightarrow \sip + \me + \me$ & 
$k_{34} = 1.88 \times 10^{-7} (1.0 + u^{0.5}) (0.376 + u)^{-1} u^{0.25} e^{-u}$ & 
$ u = 8.2 / T_{\rm e}$ & 25 \\
 & & & & \\
 \hline
 & & & & \\
35 & $\mO  + \me  \rightarrow \op  + \me + \me$ &
$k_{35} = 3.59 \times 10^{-8} (0.073 + u)^{-1} u^{0.34} e^{-u}$ & $u = 13.6 / T_{\rm e}$ & 25 \\
 & & & & \\
 \hline
 & & & & \\
36 & $\op  + \mH  \rightarrow \mO  + \Hp$ &
$ k_{36} = 4.99 \times 10^{-11} T^{0.405} +
7.54 \times 10^{-10} T^{-0.458} $ & & 26 \\
 & & & & \\
 \hline
 & & & & \\
37 & $\mO  + \Hp  \rightarrow \op  + \mH$ &
$k_{37} = [1.08 \times 10^{-11} T^{0.517} $ & & 27 \\
& & $\phantom{k_{37} = } \mbox{} + 4.00 \times 10^{-10} T^{0.00669}] \exp 
\left(-\frac{227}{T}\right)$ & & \\
 & & & & \\
 \hline
 & & & & \\
 38 & $\mO + \Hep \rightarrow \op + \He$ & 
 $k_{38} = 4.991 \times 10^{-15} \left(\frac{T}{10000}\right)^{0.3794} 
 \expf{-}{T}{1121000}$ & & 28 \\
 & & $\phantom{k_{38} = } \mbox{} + 2.780 \times 10^{-15} 
 \left(\frac{T}{10000}\right)^{-0.2163} \expf{}{T}{815800}$ & & \\
 & & & & \\
 \hline
 & & & & \\
39 & $\mC  + \Hp  \rightarrow \cp  + \mH$ & $k_{39} = 3.9 \times 10^{-16} T^{0.213}$ & & 27 \\
 & & & & \\
 \hline
  & & & & \\
40 & $\cp  + \mH  \rightarrow \mC  + \Hp$ & $k_{40} = 6.08 \times 10^{-14} 
\left(\frac{T}{10000}\right)^{1.96} \expf{-}{170000}{T}$ & & 27 \\
 & & & & \\
 \hline
 & & & & \\
 41 & $\mC + \Hep \rightarrow \cp + \He$ & 
 $k_{41} = 8.58 \times 10^{-17}  T^{0.757}$ & $T \leq 200 \: {\rm K}$ & 29 \\
& & & & \\ 
 & & $\phantom{k_{41}} = 3.25 \times 10^{-17} T^{0.968}$ & $200 < T \leq 2000 \: {\rm K}$ & \\
& & & & \\ 
 & & $\phantom{k_{41}} = 2.77 \times 10^{-19} T^{1.597}$ & $T > 2000 \: {\rm K}$ & \\
 & & & & \\ 
 \hline
 & & & & \\
42 & $\mSi + \Hp  \rightarrow \sip + \mH$ &
$k_{42} = 5.88 \times 10^{-13} T^{0.848}$ & $T \le 10^{4} \: {\rm K}$ & 30 \\
& & & & \\
& & $\phantom{k_{42}} = 1.45 \times 10^{-13} T$ & $T > 10^{4} \: {\rm K}$ & \\
 & & & & \\
\hline
 & & & & \\
 43 & $\mSi + \Hep \rightarrow \sip + \He$ & $k_{43} = 3.3 \times 10^{-9}$ & & 31 \\
 & & & & \\
 \hline
 & & & & \\
44 & $\cp  + \mSi \rightarrow \mC + \sip$ & $k_{44} = 2.1 \times 10^{-9}$ & & 31 \\
 & & & & \\
 \hline
 & & & & \\
 45 & $\sip + \Hp \rightarrow \sipp + \mH$ & 
 $k_{45} = 4.10 \times 10^{-10} \left( \frac{T}{10000} \right)^{0.24}$ & & 30 \\
 & & $\phantom{k_{45} = } \mbox{} \times \left[1.0 + 3.17 \expf{}{T}{2.39 \times 10^{6}} \right] 
 \expf{-}{3.178}{T_{\rm e}}$ & & \\
  & & & & \\
 \hline
 & & & & \\
 46 & $\sipp + \mH \rightarrow \sip + \Hp$ & $k_{46} = 1.23 \times 10^{-9} 
 \left(\frac{T}{10000}\right)^{0.24}$ & & 30 \\ 
 & & $\phantom{k_{46} =} \mbox{} \times \left[1.0 + 3.17 \expf{}{T}{2.39 \times 10^{6}} \right] $ & & \\ 
  & & & & \\
 \hline
 & & & & \\
 47 & $\sipp + \me \rightarrow \sip + \gamma$ & $k_{47, {\rm rr}} = 
 1.75 \times 10^{-12} \left( \frac{T}{10000} \right)^{-0.6346}$ & & 32 \\ 
 & & & & \\
 & & $k_{47, {\rm di}} = 2.2552 \times 10^{-11} T_{\rm e}^{-1.5} \expf{-}{2.76}{T_{\rm e}}$ & & 33 \\
& & $\phantom{k_{47, {\rm di}} = } \mbox{} + 5.6058 \times 10^{-9} T_{\rm e}^{-1.5} 
\expf{-}{10.13}{T_{\rm e}}$ & & \\
& & & & \\
\enddata
\tablerefs{1: \citet{WIS79}, 2: \citet{LAU91}, 3: \citet{RAM76}, 4: \citet{KAR79},
5: \citet{MOS70}, 6: \citet{SCH94}, 7: \citet{SAV04}, 8: \citet{STI99}, 9: \citet{MAC86},
10: \citet{MAR98}, 11: \citet{JAN87}, 12: \citet{FER92}, 13: \citet{POU78}, 
14: \citet{sav02}, 15: \citet{ger82}, 16: \citet{mlts94}, 17: \citet{s59}, 
18: \citet{hs98}, 19: \citet{ap73}, 20: \citet{z89}, 21: \citet{kldd93}, 
22: \citet{NAH97}, 23: \citet{NAH00}, 24: \citet{NAH99}, 25: \citet{VOR97},
26: \citet{STA99}, 27: \citet{STA98}, 28: \citet{z04}, 29: \citet{kd93},
30: \citet{KIN96}, 31: \citet{TEU00}, 32: \citet{n95,n96}, 33: \citet{mmcv98}}
\tablecomments{$T$ and $T_{\rm e}$ are the gas temperature in units of K 
and eV respectively. References are to the primary source of data for each 
reaction.}
\end{deluxetable}

\begin{deluxetable}{llllll}
\tablewidth{0pt}
\tabletypesize{\footnotesize}
 \tablecaption{List of the photochemical gas-phase reactions in our
chemical model. \label{tab:chem_gas_photo}}
\tablehead{No.\ & Reaction & Cross-section (${\rm cm}^{2}$) & & Reference}
\startdata
48 & $\mH + \gamma \rightarrow \Hp + \me$ & 
$\sigma_{48} = 6.3 \times 10^{-18} \left(\frac{E_{\rm th}}{E} \right)^{4} 
\exp(4 - 4\varepsilon^{-1} \arctan \: \varepsilon)$ & $E_{\rm th} = 13.6 \: {\rm eV}$ & 1 \\
& & $\phantom{\sigma_{48} = } \mbox{} \times \left[1-\exp(-2\pi/\epsilon) \right]^{-1}$
& $\varepsilon = \sqrt{\frac{E}{13.6} - 1}$ & \\
& & & & \\
49 & $\mD + \gamma \rightarrow \Dp + \me$ & $\sigma_{49} = \sigma_{48}$ & $E_{\rm th} = 13.6 \: {\rm eV}$ & 1 \\
& & & & \\
50 & $\He + \gamma \rightarrow \Hep + \me$ & 
$\sigma_{50} =  3.1451 \times 10^{-16} \left(\frac{E_{\rm th}}{E}\right)^{7/2} \times$ 
& $E_{\rm th} = 24.6 \: {\rm eV}$ & 2 \\
& & $\phantom{\sigma_{50}= } \left[1.0 - 4.7416 
\left(\frac{E_{\rm th}}{E}\right)^{1/2} + 14.82 \left(\frac{E_{\rm th}}{E}\right) \right.$  & & \\
& & $\phantom{\sigma_{50} = } \mbox{} - 30.8678 \left(\frac{E_{\rm th}}{E}\right)^{3/2}
+ 37.3584 \left(\frac{E_{\rm th}}{E}\right)^{2}$ & & \\
& & $\left. \phantom{\sigma_{50} = } \mbox{} 
-23.4585 \left(\frac{E_{\rm th}}{E}\right)^{5/2} + 5.9133 \left(\frac{E_{\rm th}}{E}\right)^{3} \right]$ & & \\
& & & & \\
51 & $\Hm + \gamma \rightarrow \mH + \me$ & 
$\sigma_{51} =  2.11 \times 10^{-16} (E - E_{\rm th})^{3/2} E^{-3}$ & $E_{\rm th} = 0.755 \: {\rm eV}$ & 3 \\
& & & & \\
52 & $\mHtp + \gamma \rightarrow \mH + \Hp$ & 
$\sigma_{52} = {\rm dex} \left[ -40.97 + 15.9795 \left(\frac{E}{E_{\rm th}}\right)
 - 3.53934 \left(\frac{E}{E_{\rm th}}\right)^{2} \right.$ & $E_{\rm th} = 2.65 \: {\rm eV}$ & 4 \\
& & $\left. \phantom{\sigma_{52}=} \mbox{} + 0.2581155  \left(\frac{E}{E_{\rm th}}\right)^{3} \right]$ 
& $2.65 < E < 11.27 \: {\rm eV}$ & \\
& & & & \\
& & $\phantom{\sigma_{52}} = {\rm dex} \left[ -30.26 + 7.3935 \left(\frac{E}{E_{\rm th}}\right)
 - 1.29214 \left(\frac{E}{E_{\rm th}}\right)^{2} \right. $ & $11.27 < E < 21.0 \: {\rm eV}$ & \\ 
& & $\left. \phantom{\sigma_{52}=} \mbox{} + 6.5785 \times 10^{-2} \left(\frac{E}{E_{\rm th}}\right)^{3} \right]$ & & \\ 
& & & & \\
53 & $\mHt + \gamma \rightarrow \mH + \mH$ & See \S\ref{h2_photodiss}  &  & 5 \\
& & & & \\
54 & $\mHt + \gamma \rightarrow \mHtp + \me$ & 
$\sigma_{54} =  9.560 \times 10^{-17} \left(\frac{E}{E_{\rm th}}\right) - 9.4 \times 10^{-17}$ &
$ E_{\rm th} = 15.4 \: {\rm eV}$ & 6 \\
& & & $15.4 < E < 16.5 \: {\rm eV}$ & \\
& & & & \\
& & $\phantom{\sigma_{54}} =  2.16 \times 10^{-17}  \left(\frac{E}{E_{\rm th}}\right) 
- 1.48 \times 10^{-17}$ & $16.5 < E < 17.7 \: {\rm eV}$ & \\
& & & & \\
& & $\phantom{\sigma_{54}} = 1.51 \times 10^{-17} \left(\frac{E}{E_{\rm th}}\right)^{-2.71}$ &
$17.7 < E < 30.0 \: {\rm eV}$ & \\
& & & & \\
55 & $\hd + \gamma \rightarrow \mH + \mD$ & See \S\ref{h2_photodiss}  & & 7 \\ 
& & & & \\
56 & $\mC + \gamma \rightarrow  \cp  + \me$ & 
$\sigma_{56} = 5.027 \times 10^{-16} F(x,y,y_{w},y_{a},P)$ & $E_{\rm th} = 11.26 \: {\rm eV}$ & 8 \\
& & & $x = \frac{E}{2.144} - 1.133$ & \\
& & & $y = \sqrt{x^{2} + 1.607^2}$ & \\
& & & $y_{w} = 0.09157$ & \\
& & & $y_{a} = 62.16$ & \\
& & & $P = 5.101$ & \\
& & & & \\
57 & $\mO + \gamma \rightarrow \op + \me$ & 
$\sigma_{57} = 1.745 \times 10^{-15} F(x,y,y_{w},y_{a},P)$ & $E_{\rm th} = 13.62 \: {\rm eV}$ & 8 \\
& & & $x = \frac{E}{1.240} - 8.698$ & \\
& & & $y = \sqrt{x^{2} + 0.1271^2}$ & \\
& & & $y_{w} = 0.07589$ & \\
& & & $y_{a} = 3.784$ & \\
& & & $P = 17.64$ & \\
& & & & \\
& & & & \\
58 & $\mSi + \gamma \rightarrow \sip + \me$ & 
$\sigma_{58} = 2.506 \times 10^{-17} F(x,y,y_{w},y_{a},P)$ & $E_{\rm th} = 8.152 \: {\rm eV}$ & 8 \\
& & & $x = \frac{E}{23.17} - 1.672 \times 10^{-5}$ & \\
& & & $y = \sqrt{x^{2} + 0.4207^2}$ & \\
& & & $y_{w} = 0.2837$ & \\
& & & $y_{a} = 20.57$ & \\
& & & $P = 3.546$ & \\
& & & & \\
59 & $\sip + \gamma \rightarrow \sipp + \me$ & 
$\sigma_{59} = 4.140 \times 10^{-18} F(x,y,y_{w},y_{a},P)$ & $E_{\rm th} = 16.35 \: {\rm eV}$ & 8 \\
& & & $x = \frac{E}{2.556} - 6.634$ & \\
& & & $y = \sqrt{x^{2} + 0.1272^2}$ & \\
& & & $y_{w} = 1.570$ & \\
& & & $y_{a} = 13.37$ & \\
& & & $P = 11.91$ & \\
\enddata
\tablerefs{1: \citet{os89}, 2: \citet{ysd98}, 3: \citet{dj72,SHA87}, 4: \citet{DUN68}, 5: \citet{DRA96}, 
6: \citet{or78,wam02}, 7: \citet{ar06}, 8: \citet{VER96}} 
\tablecomments{References are to the primary source of data for each reaction.
$E$ is the photon energy in eV and $E_{\rm th}$ is the energy threshold in eV.
The fitting function $F$ used in the tabulated cross-sections for reactions 56--59
is from \citet{VER96} and is given
by $F = [(x-1)^{2} + y_{w}^{2}] y^{0.5P-5.5} (1+\sqrt{y/y_{a}})^{-P}$.
Photodissociation of $\mHt$ and $\hd$ occurs via absorption into a large number
of discrete spectral lines and so no simple cross-section can be given for these 
processes; see \S\ref{h2_photodiss}  for more details}
\end{deluxetable}

\begin{deluxetable}{lllc}
\tablewidth{0pt}
\tablecaption{List of the grain surface reactions included in our 
 chemical model. \label{tab:chem_grain}}
 \tablehead{No.\  & Reaction & Rate coefficient $({\rm cm}^{3} \: {\rm s}^{-1})$ &
Ref.\ }
 \startdata
& & & \\
60 & $\mH + \mH \rightarrow \mHt$ &
$k_{60} = 3.0 \times 10^{-18} T^{0.5} ({\cal D}/{\cal D_{\odot}}) 
[1.0 + 4\times 10^{-2}(T + T_{\rm gr})^{0.5}$ & 1 \\
& & $ \phantom{k_{60} =} \mbox{} + 2 \times 10^{-3} T + 8 \times 10^{-6} T^{2}]^{-1} \left[1.0 + 10^{4} 
\exp \left(-\frac{600}{T_{\rm gr}}\right) \right]^{-1}$ & \\
& & & \\
61 & $\Hp + \me \rightarrow \mH$  & 
$k_{61} = 1.225 \times 10^{-13} ({\cal D}/{\cal D_{\odot}}) [1.0 + 8.074 \times 10^{-6} \psi^{1.378}$ & 2 \\
 & & $\phantom{k_{61} =} (1.0 + 5.087 \times 10^{2} T^{0.01586} \psi^{-0.4723 - 
 1.102 \times 10^{-5}\ln T})]^{-1}$ & \\
& & & \\
62 & $\Dp + \me \rightarrow \mD$ & $k_{62} = \frac{1}{\sqrt{2}} k_{61}$ & 3 \\
& & & \\
63 & $\Hep + \me \rightarrow \He$ & 
$k_{63} = 5.572 \times 10^{-14} ({\cal D}/{\cal D_{\odot}}) [1.0 + 3.185 \times 10^{-7} \psi^{1.512}$ & 2 \\
 & & $\phantom{k_{63} =} (1.0 + 5.115 \times 10^{3} T^{3.903 \times 10^{-7}} \psi^{-0.4956 - 
 5.494 \times 10^{-7}\ln T})]^{-1}$ & \\
& & & \\
64 & $\cp + \me \rightarrow \mC$  &
$k_{64} = 4.558 \times 10^{-13} ({\cal D}/{\cal D_{\odot}}) [1.0 + 6.089 \times 10^{-3} \psi^{1.128} $ & 2 \\
 & & $\phantom{k_{64} =} (1.0 + 4.331 \times 10^{2} T^{0.04845} \psi^{-0.8120 - 
 1.333 \times 10^{-4}\ln T})]^{-1}$ & \\
& & & \\
65 & $\op + \me \rightarrow \mO$ & $k_{65} = \frac{1}{4} k_{61}$ & 3 \\
& & & \\
66 & $\sip + \me \rightarrow \mSi$  &
$k_{66} = 2.166 \times 10^{-14} ({\cal D}/{\cal D_{\odot}}) [1.0 + 5.678 \times 10^{-8} \psi^{1.874} $ & 2 \\
& & $\phantom{k_{65}=} (1.0 + 4.375 \times 10^{4} T^{1.635\times10^{-6}} \psi^{-0.8964 - 
7.538 \times 10^{-5}\ln T})]^{-1}$ & \\
& & & \\
\enddata
\tablecomments{${\cal D}$ is the dust-to-gas ratio and ${\cal D}_{\odot}$ is the dust-to-gas ratio
in the local ISM. We generally assume that ${\cal D}/{\cal D}_{\odot} \equiv {\rm Z}/{\rm Z_{\odot}}$.
$T$ and $T_{\rm gr}$ are the gas and grain temperatures, respectively. 
The parameter $\psi$ in the grain recombination rates is given by $\psi = G\sqrt{T} / n_{\rm e}$,
where $G \simeq 0.01 J_{21}$ is a measure of the radiation energy density between 
$6 \: {\rm eV}$  and $13.6 \: {\rm eV}$ relative to the \citet{habing68} field.}
\tablerefs{1: \citet{HOL79}; 2: \citet{WEI01a}; 3: This work, but based on \citet{WEI01a}}
\end{deluxetable}

\begin{deluxetable}{clcc}
\tablecaption{List of cosmic ray ionization processes included in our chemical model
\label{tab:cosmic}}
\tablewidth{0pt}
\tablehead{No.\ & Reaction & $\zeta_{\rm i}/\zeta_{\mH}$ & Ref.\ }
\startdata
67 & $\mH + {\rm c.r.} \rightarrow \Hp + \me$ & 1.0 & \citet{TEU00} \\
68 & $\mD + {\rm c.r.} \rightarrow \Dp + \me$ & 1.0 & \citet{TEU00} \\
69 & $\He  + {\rm c.r.} \rightarrow \Hep + \me$ & 1.09 & \citet{TEU00} \\
70 & $\mHt + {\rm c.r.} \rightarrow \mHtp + \me$ & 2.0 & \citet{TEU00} \\
71 & $\mC  + {\rm c.r.} \rightarrow \cp + \me$ & 3.83 & \citet{TEU00} \\
72 & $\mO  + {\rm c.r.} \rightarrow \op + \me$ & 5.67 & \citet{TEU00} \\
73 & $\mSi + {\rm c.r.} \rightarrow \sip + \me$ & 6.5 & \citet{lotz67,lang78} \\
74 & $\sip + {\rm c.r.} \rightarrow \sipp + \me$ & 2.5 & \citet{lotz67,lang78} \\
\enddata
\tablecomments{We list here the ratio of the various rates to the rate of 
process 67, the cosmic ray ionization of atomic hydrogen, $\zeta_{\mH}$,
which we treat as an adjustable parameter in our models. Rates for cosmic
ray ionization of $\mSi$ and $\sip$ were calculated following the prescription
in \citet{lang78} and using data from \citet{lotz67} under the assumption that
the effective number of outer shell electrons for $\mSi$ and $\sip$ in the
high energy limit is the same as that for $\mC$ and $\cp$.}
\end{deluxetable}

\begin{deluxetable}{cclllll}
\tablecaption{Atomic data for the fine structure transitions included in our thermal model  \label{fs_data} }
\tablewidth{0pt}
\tablehead{\colhead{Coolant} & \colhead{Transition} 
& \colhead{$g_{j}$} & \colhead{$g_{i}$} 
& \colhead{$\lambda_{ji} (\mu{\rm m})$} & \colhead{$E_{ji} / k \: ({\rm K})$} & 
\colhead{$A_{ji} \: ({\rm s^{-1}})$}}
\startdata
$\mC$ & $1 \rightarrow 0$ & 3 & 1 & 609.2 & 24 & $7.9 \times 10^{-8}$ \\
$\mC$ & $2 \rightarrow 0$ & 5 & 1 & 229.9 & 63 & $2.1 \times 10^{-14}$ \\
$\mC$ & $2 \rightarrow 1$ & 5 & 3 & 369.0 & 39 & $2.7 \times 10^{-7}$ \\
$\mO$ & $1 \rightarrow 0$ & 3 & 5 & 63.1& 230  & $8.9 \times 10^{-5}$ \\
$\mO$ & $2 \rightarrow 0$ & 1 & 5 & 44.2 & 330  & $1.3 \times 10^{-10}$ \\
$\mO$ & $2 \rightarrow 1$ & 1 & 3 & 145.6 & 98 & $1.8 \times 10^{-5}$  \\
$\mSi$ & $1 \rightarrow 0$ & 3 & 1 & 129.6 & 110 & $8.4 \times 10^{-6}$  \\ 
$\mSi$ & $2 \rightarrow 0$ & 5 & 1 & 44.8 & 320 & $2.4 \times 10^{-10}$ \\
$\mSi$ & $2 \rightarrow 1$ & 5 & 3 & 68.4 & 210 & $4.2 \times 10^{-5}$  \\
$\cp$  & $1 \rightarrow 0$ & 4 & 2 & 157.7 & 92 & $2.3 \times 10^{-6}$  \\
$\sip$ & $1 \rightarrow 0$ & 4 & 2 & 34.8 & 410  & $2.2 \times 10^{-4}$  \\
\enddata
\end{deluxetable}

\begin{deluxetable}{cclcc}
\tablecaption{Collisional de-excitation rates for atomic fine-structure coolants \label{fs_coll_rates}}
\tablewidth{0pt}
\tabletypesize{\footnotesize}
\tablehead{\colhead{Coolant} & \colhead{Collider} & 
\colhead{De-excitation rates (${\rm cm^{3}} \: {\rm s^{-1}}$)} &
\colhead{Temperature range (K)} & \colhead{Refs.}}
\startdata
& & & & \\
$\mC$ & o-$\mHt$ & $q_{10} = 8.7 \times 10^{-11} - 6.6 \times 10^{-11}  \expf{-}{T}{218.3}$ & &  \\
& & $\phantom{q_{10}} \mbox{} + 6.6 \times 10^{-11} \expf{-}{2T}{218.3}$ & &  1 \\
 & & $q_{20} = 1.2 \times 10^{-10} - 6.1 \times 10^{-11} \expf{-}{T}{387.3}$ & & 1 \\
 & & $q_{21} = 2.9 \times 10^{-10} - 1.9 \times 10^{-10} \expf{-}{T}{348.9}$ & & 1 \\
& & & & \\
$\mC$ & p-$\mHt$ & $q_{10} = 7.9 \times 10^{-11} - 8.7 \times 10^{-11}  \expf{-}{T}{126.4}$ & & \\
& & $\phantom{q_{10}} \mbox{} + 1.3 \times 10^{-10} \expf{-}{2T}{126.4}$ & & 1  \\
& & $q_{20} = 1.1 \times 10^{-10} - 8.6 \times 10^{-11} \expf{-}{T}{223.0} $ & &  \\
& & $\phantom{q_{20}} \mbox{} + 8.7 \times 10^{-11} \expf{-}{2T}{223.0} $ & & 1 \\
& & $q_{21} = 2.7 \times 10^{-10} - 2.6 \times 10^{-10} \expf{-}{T}{250.7} $ & &  \\
& & $\phantom{q_{21}} \mbox{} + 1.8 \times 10^{-10} \expf{-}{2T}{250.7}$ & & 1 \\
& & & & \\
$\mC$ & $\mH$ & $q_{10} = 1.6 \times 10^{-10} T_{2}^{0.14}$ & & 2 \\
& & $q_{20} = 9.2 \times 10^{-11} T_{2}^{0.26}$ & & 2 \\
& & $q_{21} = 2.9 \times 10^{-10} T_{2}^{0.26}$ & & 2 \\
& & & & \\
$\mC$ & $\Hp$ & $q_{10} = (9.6 \times 10^{-11} - 1.8 \times 10^{-14} T + 1.9 \times 10^{-18} T^{2}) 
T^{0.45}$ & $T \le 5000 $ & 3 \\
& & $\phantom{q_{10}} = 8.9 \times 10^{-10} T^{0.117}$ & $T > 5000 $ & 3 \\
& & & & \\
& & $q_{20} = (3.1 \times 10^{-12} - 6.0 \times 10^{-16} T + 3.9 \times 10^{-20} T^{2}) T$ & 
$T \le 5000 $ & 3 \\
& & $\phantom{q_{20}} = 2.3 \times 10^{-9} T^{0.0965}$ & $T > 5000 $ & 3 \\
& & & & \\
& & $q_{21} = (1.0 \times 10^{-10} - 2.2 \times 10^{-14} T + 1.7 \times 10^{-18} T^{2}) T^{0.70}$ 
& $T \le 5000 $ & 3 \\
& & $\phantom{q_{21}} = 9.2 \times 10^{-9} T^{0.0535}$ & $T > 5000 $ & 3 \\
& & & & \\
$\mC$ & $\me$ & $q_{10} = 2.88 \times 10^{-6} T^{-0.5} \exp [ -9.25141 
- 7.73782 \times 10^{-1} \ln{T}$ & & \\
& & $\phantom{q_{10}} \mbox{} + 3.61184 \times 10^{-1} (\ln{T})^{2}
 - 1.50892 \times 10^{-2} (\ln{T})^{3}$ & & \\
& & $\phantom{q_{10}} \mbox{} - 6.56325 \times 10^{-4} (\ln{T})^{4}]$ & $T \le 1000 $ & 4 \\
& & & & \\
& & $\phantom{q_{10}} = 2.88 \times 10^{-6} T^{-0.5} \exp [ -4.44600 \times 10^{2} 
- 2.27913 \times 10^{2} \ln{T}$ & & \\
& & $\phantom{q_{10}} \mbox{} + 4.2595 \times 10^{1} (\ln{T})^{2} 
- 3.47620 \times 10^{0} (\ln{T})^{3}$ & & \\
& & $\phantom{q_{10}} \mbox{} + 1.0508 \times 10^{-1} (\ln{T})^{4}]$ & $T > 1000 $ & 4 \\
& & & & \\ 
& & $q_{20} = 1.73 \times 10^{-6} T^{-0.5} \exp [ -7.69735 - 1.30743 \ln{T}$ & & \\
& & $\phantom{q_{20}} \mbox{} + 0.697638 (\ln{T})^{2} - 0.111338 (\ln{T})^{3}$ & & \\
& & $\phantom{q_{20}} \mbox{} + 0.705277 \times 10^{-2} (\ln{T})^4 ]$ & $T \le 1000 $ & 4 \\
& & & & \\
& & $\phantom{q_{20}} = 1.73 \times 10^{-6} T^{-0.5} \exp [3.50609 \times 10^{2} 
- 1.87474 \times 10^{2} \ln{T}$ & & \\
& & $\phantom{q_{20}} \mbox{} + 3.61803 \times 10^{1} (\ln{T})^{2} 
- 3.03283 \times 10^{0} (\ln{T})^{3}$ & & \\
& & $\phantom{q_{20}} \mbox{} + 9.38138 \times 10^{-2} (\ln{T})^{4} ]$ & $T > 1000 $ & 4 \\
& & & & \\
& & $q_{21} = 1.73 \times 10^{-6} T^{-0.5} \exp [ -7.4387 - 0.57443  \ln{T}$ & & \\
& & $\phantom{q_{21}} \mbox{} + 0.358264 (\ln{T})^{2}  - 4.18166 \times 10^{-2} (\ln{T})^{3}$ & & \\
& & $\phantom{q_{21}} \mbox{} + 2.35272 \times 10^{-3} (\ln{T})^{4} ]$ & $T \le 1000 $ & 4 \\
& & & & \\
& & $\phantom{q_{21}} = 1.73 \times 10^{-6} T^{-0.5} \exp [ 3.86186 \times 10^{2} 
- 2.02192 \times 10^{2} \ln{T}$ & & \\
& & $\phantom{q_{21}} \mbox{} + 3.85049 \times 10^{1} (\ln{T})^{2} 
- 3.19268 \times 10^{0} (\ln{T})^{3}$ & & \\
& & $\phantom{q_{21}} \mbox{} + 9.78573 \times 10^{-2} (\ln{T})^{4}]$ & $T > 1000 $ & 4 \\
& & & & \\
\hline
& & & & \\
$\mO$ & o-$\mHt$ & $q_{10} = 2.7 \times 10^{-11} T^{0.362} $ & & 5 \\
& & $q_{20} = 5.49 \times 10^{-11} T^{0.317} $ & & 5 \\
& & $q_{21} = 2.74 \times 10^{-14} T^{1.060} $ & & 5 \\
& & & & \\
$\mO$ & p-$\mHt$ & $q_{10} = 3.46 \times 10^{-11} T^{0.316} $ & & 5 \\
& & $q_{20} = 7.07 \times 10^{-11} T^{0.268} $ & & 5 \\
& & $q_{21} = 3.33 \times 10^{-15} T^{1.360} $ & & 5 \\
& & & & \\
$\mO$ & $\mH$ & $q_{10} = 9.2 \times 10^{-11} T_{2}^{0.67} $ & & 5 \\
& & $q_{20} = 4.3 \times 10^{-11} T_{2}^{0.80} $ & & 5 \\
& & $q_{21} = 1.1 \times 10^{-10} T_{2}^{0.44} $ & & 5 \\
& & & & \\
$\mO$ & $\Hp$ & $q_{10} = 6.38 \times 10^{-11} T^{0.40}$ & $T \le 194 $ & 6 \\
& & $\phantom{q_{10}} = 7.75 \times 10^{-12} T^{0.80}$ & $194 < T  \le 3686 $ & \\
& & $\phantom{q_{10}} = 2.65 \times 10^{-10} T^{0.37}$ & $T > 3686 $ & \\
& & & & \\
& & $q_{20} = 6.10 \times 10^{-13} T^{1.10}$ & $T \le 511 $ & 6 \\
& & $\phantom{q_{20}} = 2.12 \times 10^{-12} T^{0.90}$ & $511 < T  \le 7510 $ & \\
& & $\phantom{q_{20}} = 4.49 \times 10^{-10} T^{0.30}$ & $T > 7510 $ & \\
& & & & \\
& & $q_{21} = 2.03 \times 10^{-11} T^{0.56}$ & $T \le 2090 $ & 6 \\
& & $\phantom{q_{21}} = 3.43 \times 10^{-10} T^{0.19}$ & $T > 2090 $ & \\
& & & & \\
$\mO$ & $\me$ & $q_{10} = 5.12 \times 10^{-10} T^{-0.075} $ & & 7 \\
& & $q_{20} = 4.86 \times 10^{-10} T^{-0.026} $ & & 7 \\
& & $q_{21} = 1.08 \times 10^{-14} T^{0.926} $ & & 7 \\
& & & & \\
\hline
& & & & \\
$\mSi$ & $\mH$ & $q_{10} =  3.5 \times 10^{-10} T_{2}^{-0.03}$ & & 2 \\
& & $q_{20} = 1.7 \times 10^{-11} T_{2}^{0.17} $ & & 2 \\
& & $q_{21} = 5.0 \times 10^{-10} T_{2}^{0.17} $ & & 2 \\
& & & & \\
$\mSi$ & $\Hp$ & $q_{10} = 7.2 \times 10^{-9} $ & & 2 \\
& & $q_{20} = 7.2 \times 10^{-9} $ & & 2 \\
& & $q_{21} = 2.2 \times 10^{-8} $ & & 2 \\
& & & & \\
\hline
& & & & \\
$\cp$ & o-$\mHt$ & $q_{10} = 4.7 \times 10^{-10} + 4.6 \times 10^{-13} T$  & $T \le 250 $ & 8, 9 \\
& & $\phantom{q_{10}} = 5.85 \times 10^{-10} T^{0.07}$ & $T > 250 $ & \\
& & & & \\
$\cp$ & p-$\mHt$ & $q_{10} = 2.5 \times 10^{-10} T^{0.12}$ & $T \le 250 $ & 8, 9 \\
& & $\phantom{q_{10}} = 4.85 \times 10^{-10} T^{0.07}$ & $T > 250 $ & \\
& & & & \\
$\cp$ & $\mH$ & $q_{10} = 8.0 \times 10^{-10} T_{2}^{0.07}$ & $T \le 2000 $ & 2, 10 \\
& & $\phantom{q_{10}} = 3.1 \times 10^{-10} T_{2}^{0.385}$ & $T > 2000 $ & \\
& & & & \\
$\cp$ & $\me$ & $q_{10} = 3.86 \times 10^{-7} T_{2}^{-0.5}$ & $T \le 2000 $ & 11 \\
& & $\phantom{q_{10}} = 2.43 \times 10^{-7} T_{2}^{-0.345}$ & $T > 2000 $ & \\
& & & & \\
\hline
& & & & \\
$\sip$ & $\mH$ & $q_{10} = 4.95 \times 10^{-10} T_{2}^{0.24}$ & & 12 \\
& & & & \\
$\sip$ & $\me$ & $q_{10} = 1.2 \times 10^{-6} T_{2}^{-0.5}$ & & 13 \\
& & & & \\
\enddata
\tablecomments{o-$\mHt$ and p-$\mHt$ denote ortho-$\mHt$ and para-$\mHt$
respectively. $T$ is the gas temperature (in Kelvin) and $T_{2} = 10^{-2} T$.}
\tablerefs{1: \citet{SCH91}; 2: \citet{HOL89}; 3: \citet{RLB90}; 4: \citet{JOH87}; 
5: Flower, private communication; 6:  \citet{PEQ90,PEQ96}; 7: \citet{BEL98}; 
8: \citet{FLO77}; 9: assumed to have the same scaling with $T$ as the low temperature
$\mH$ rate for temperatures above the range of the \citet{FLO77} fit; 
10: \citet{KEE86}; 11: \citet{WIL02};  12: \citet{ROU90}; 
13: \citet{DUF91}, extrapolated to $T < 4000 \: {\rm K}$ assuming constant 
collision strength}
\end{deluxetable}

\begin{deluxetable}{llc}
\tablecaption{Other processes included in our thermal model. \label{cool_other}}
\tablewidth{0pt}
\tabletypesize{\small}
\tablehead{
\colhead{Process} & \colhead{Rate (${\rm erg \: cm^{-3}} \: {\rm s^{-1}}$)} & \colhead{Ref.} }
\startdata
{\bf Cooling:} & & \\
& & \\
H excitation  & $\Lambda = 7.5 \times 10^{-19} \left(1.0 + \sqrt{T/10^{5}}\right)^{-1}
 \expf{-}{118348}{T} n_{\rm e} n_{\mH}$ & 1 \\
& & \\
He excitation ($1^1$S state) & $\Lambda = 1.1 \times 10^{-19} T^{0.082} \expf{-}{230000}{T} n_{\rm e} n_{\He}$ & 2 \\
& & \\
He excitation ($2^3$S state) & $\Lambda = 9.1 \times 10^{-27} T^{-0.1687} \left(1.0 + \sqrt{T/10^{5}}\right)^{-1}
\expf{-}{13179}{T} n_{\rm e}^{2} n_{\Hep}$  & 1 \\
& & \\
$\Hep$ excitation &$\Lambda = 5.54 \times 10^{-17} T^{-0.397} \left(1.0 + \sqrt{T/10^{5}}\right)^{-1}
 \expf{-}{473638}{T} n_{\rm e} n_{\Hep}$ & 1 \\
& & \\
$\mH$ collisional ionization  & $\Lambda = 2.179 \times 10^{-11} k_{11} n_{\rm e} n_{\mH}$ & 3 \\
& & \\
$\He$ collisional ionization & $\Lambda = 3.94 \times 10^{-11} k_{24} n_{\rm e} n_{\He}$ & 3 \\
& & \\
Compton cooling & $\Lambda = 1.017 \times 10^{-37} T_{\rm CMB}^{4} \left(T - T_{\rm CMB}\right) n_{\rm e}$ & 1 \\
& & \\
Bremsstrahlung & $\Lambda = 1.426 \times 10^{-27} Z_{i}^{2} T^{1/2} g_{\rm ff}(Z_{i}, T) n_{\rm e} n_{i}$   & 4 \\
& & \\
& $g_{\rm ff} = 0.79464 + 0.1243 \log \left(T/Z_{i}^{2}\right)  \hspace{.5in} (T/Z_{i}^{2}) < 320000 \: {\rm K} $ & \\
& $\phantom{g_{\rm ff}} =  2.13164 - 0.1240 \log \left(T/Z_{i}^{2}\right) \hspace{.5in} (T/Z_{i}^{2}) > 320000 \: {\rm K} $ & \\
& & \\
$\Hp$ recombination (radiative) & $\Lambda = 1.38 \times 10^{-16} T k_{13} n_{e} n_{\Hp}$ & 5 \\
& & \\
$\Hep$ recombination (radiative) & $\Lambda = 1.38 \times 10^{-16} T k_{25, \rm rr} n_{e} n_{\Hep}$  & 6 \\
& & \\
$\Hep$ recombination (dielectronic) &  $\Lambda = 6.54 \times 10^{-11} k_{25, \rm di} n_{e} n_{\Hep}$ & 7 \\
& & \\
Grain surface recombination & $\Lambda = 2.33 \times 10^{-30} T^{0.94} \tilde{\psi}^{0.74 / T^{0.068}} 
\left(\frac{{\rm Z}}{{\rm Z_{\odot}}}\right) n_{\rm e} n$ & 8 \\
& & \\
$\mHt$ rovibrational lines & See \S\ref{cool:other} & 9 \\
& & \\
$\hd$ rovibrational lines & See \S\ref{cool:other} & 10 \\
& & \\
$\mHt$ collisional dissociation & $\Lambda =  7.2 \times 10^{-12} \left(k_{9} n_{\mH} + k_{10} n_{\mHt} \right) n_{\mHt}$ & 11 \\
& & \\
Gas-grain energy transfer & $\Lambda = 3.8 \times 10^{-33} T^{1/2} (T - T_{\rm gr})
\left[1.0 - 0.8 \expf{-}{75}{T}\right] \left(\frac{{\rm Z}}{{\rm Z_{\odot}}}\right) n^{2}$ & 12 \\
& & \\
& & \\
{\bf Heating:} & & \\
& & \\
Photoelectric effect & $\Gamma = 1.3 \times 10^{-24} \epsilon \, G \left(\frac{{\rm Z}}{{\rm Z_{\odot}}}\right) n$& 13 \\
& & \\
& $\epsilon =  \frac{4.9 \times 10^{-2}}{1.0 + 4.0 \times 10^{-3} \tilde{\psi}^{0.73}} + \frac{3.7 \times 10^{-2} 
(T/10000)^{0.7}}{1.0 + 2.0 \times 10^{-4} \tilde{\psi}}$ & \\
& & \\
$\mHt$ photodissociation & $\Gamma = 6.4 \times 10^{-13} R_{\rm diss} n_{\mHt}$ & 14 \\ 
& & \\
UV pumping of $\mHt$ & $\Gamma = 2.7 \times 10^{-11} R_{\rm diss} n_{\mHt} \left(\frac{n}{n + n_{\rm cr}}\right)$  & 15 \\
& & \\
H photoionization & Dependent on incident spectrum; see \S\ref{photochem} & 16 \\
& & \\
He photoionization & Dependent on incident spectrum; see \S\ref{photochem}  & 17 \\
& & \\
Gas-phase $\mHt$ formation  & $\Gamma = \left[2.93 \times 10^{-12} k_{2} n_{\Hm} +
5.65 \times 10^{-12} k_{4} n_{\mHtp}\right] n_{\mH} \left(\frac{n}{n + n_{\rm cr}}\right)$ & 18 \\
& & \\
$\mHt$ formation on dust grains & $\Gamma = 7.16 \times 10^{-12} k_{60} \, n \, n_{\mH} 
\left(\frac{n}{n + n_{\rm cr}}\right)$ & 19 \\
& & \\
Cosmic-ray ionization & $\Gamma = 3.2 \times 10^{-11} \zeta_{\rm tot} \, n$ & 20 \\
\enddata
\tablerefs{1: \citet{CEN92}, 2: \citet{BBFT00}, 3: \citet{JAN87}, 4: \citet{SHA87}
5: \citet{FER92}, 6: \citet{hs98}, 7: \citet{ap73}, 8: \citet{WOL03}, 9: \citet{BOU99}, 
10: \citet{lna05}, 11: \citet{MAC86, MAR98}, 12: \citet{HOL89}, 13: \citet{BAK94,WOL95}, 
14: \citet{BLA77}, 15: \citet{BUR90}, 16: \citet{os89}, 17: \citet{ysd98}, 18: \citet{LAU91,KAR79}, 
19: \citet{HOL79}, 20: \citet{gl78}}
\tablecomments{$Z_{i}$ and $n_{i}$ are the ion charge and
number density of ion $i$. The parameter $\tilde{\psi}$ is given by 
$\tilde{\psi} =  G \sqrt{T} / 0.5 n_{\rm e}$, where $G  \simeq 0.01 J_{21}$ is a 
measure of the radiation energy density between 
$6 \: {\rm eV}$  and $13.6 \: {\rm eV}$ relative to the \citet{habing68} field. 
$R_{\rm diss}$ is the photodissociation rate, calculated as discussed in \S\ref{h2_photodiss}. 
$\zeta_{\rm tot}$ is the total cosmic-ray ionization rate (i.e.\ the sum of the rates for the
various different species, weighted by their fractional abundances:
$\zeta_{\rm tot} = \sum_{i} x_{i} \zeta_{i}$). Finally, note that our treatment of recombination 
cooling here is approximate, but that it should be accurate 
enough for most purposes.}
\end{deluxetable}

\end{document}